\documentclass[useAMS,usenatbib,usegraphicx]{mn2e}

\usepackage[british]{babel}
\usepackage{times}
\usepackage{latexsym,amssymb}
\usepackage[fleqn]{amsmath}
\usepackage{graphicx}
\usepackage{natbib}
\usepackage{journalnames}
\usepackage{color}

\usepackage[T1]{fontenc}
\usepackage{aecompl}

\usepackage{caption}
\usepackage{comment}
\usepackage{lipsum}
\usepackage{array}
\usepackage{dcolumn}
\usepackage{rotating}

\title[Graph-based interpretation of the Molecular Interstellar Medium Segmentation]{Graph-based interpretation of the Molecular Interstellar Medium Segmentation}

\author[D. Colombo et al.]{%
Dario Colombo$^1$\thanks{E-mail: dcolombo@ualberta.ca},
Erik Rosolowsky$^1$, Adam Ginsburg$^2$, Ana Duarte-Cabral$^3$, Annie Hughes$^{4,5,6}$\\
$^1$Department of Physics, University of Alberta, 4-181 CCIS, Edmonton, AB T6G 2E1, Canada\\
$^2$European Southern Observatory, Karl-Schwarzschild-Strasse 2, 85748, Garching bei Munchen, Germany\\
$^3$School of Physics and Astronomy, University of Exeter, Stocker Road, Exeter EX4 4QL, UK\\
$^4$ CNRS, IRAP, 9 Av. colonel Roche, BP 44346, F-31028 Toulouse cedex 4, France\\
$^5$ Universit\'{e} de Toulouse, UPS-OMP, IRAP, F-31028 Toulouse cedex 4, France\\
$^6$ Max Planck Institute for Astronomy, K\"onigstuhl 17, 69117 Heidelberg, Germany\\
}
\pagerange{\pageref{firstpage}--\pageref{lastpage}} \pubyear{2015}

\begin{document}

\date{Draft \today}

\label{firstpage}

\maketitle

\begin{abstract}
We present a generalization of the Giant Molecular Cloud (GMC) identification problem based on cluster analysis. The method we designed, \texttt{SCIMES} (Spectral Clustering for Interstellar Molecular Emission Segmentation) considers the dendrogram of emission in the broader framework of graph theory and utilizes spectral clustering to find discrete regions with similar emission properties. For Galactic molecular cloud structures, we show that the characteristic volume and/or integrated CO luminosity are useful criteria to define the clustering, yielding emission structures that closely reproduce ``by-eye'' identification results. \texttt{SCIMES} performs best on well-resolved, high-resolution data, making it complementary to other available algorithms. Using $^{12}$CO(1-0) data for the Orion-Monoceros complex, we demonstrate that \texttt{SCIMES} provides robust results against changes of the dendrogram-construction parameters, noise realizations and degraded resolution. By comparing \texttt{SCIMES} with other cloud decomposition approaches, we show that our method is able to identify all canonical clouds of the Orion-Monoceros region, avoiding the over-division within high resolution survey data that represents a common limitation of several decomposition algorithms. The Orion-Monoceros objects exhibit hierarchies and size-line width relationships typical to the turbulent gas in molecular clouds, although ``the Scissors'' region deviates from this common description. \texttt{SCIMES} represents a significant step forward in moving away from pixel-based cloud segmentation towards a more physical-oriented approach, where virtually all properties of the ISM can be used for the segmentation of discrete objects.      
\end{abstract}


%=====================================================================
\section{Introduction}
\label{S:intro}
%=====================================================================
 
The formation of stars is strongly connected to the molecular phase of the interstellar medium (ISM; e.g., \citealt{bigiel08}, \citealt{schruba11}). Since the molecular phase is naturally clumpy on different scales (\citealt{leroy13}), it has become customary to divide the emission into isolated, independent entities named Giant Molecular Clouds, a practice which began with the earliest surveys (e.g., \citealt{scoville75}).  The first studies of the GMCs in the Galaxy defined the standard paradigm of these objects, utilized also today to define new surveys of nearby galaxies. From the seminal paper of \cite{solomon79}, GMCs possess a H$_{2}$ mass between $10^{5}-10^{6.5}$\,M$_{\odot}$, a mean H$_{2}$ density of 300\,cm$^{-3}$ and an average size of 40\,pc. From the number density of the GMCs in the Galactic Ring, the authors also calculated that the Galaxy should contain $\sim4000$ GMCs encompassing $\sim85\%$ of the Galactic molecular gas budget. Later, more comprehensive studies of the GMCs (e.g. \citealt{larson81}; \citealt{solomon87}), defined scaling relations between their properties that laid the foundations for a better understanding of the physics of the molecular ISM. In particular, GMCs appear gravitationally bound, with a roughly constant mass surface density, and with supersonic velocity dispersions proportional to the square root of their sizes. Multi-tracer observations have shown that the structure of the GMCs is essentially hierarchical: small scale dense clumps are always contained in larger, lower density gas envelopes (see, e.g. \citealt{rosolowsky08}, and references therein). Taken together, these evidences suggest that GMCs are sustained against global collapse by turbulent motions (e.g., \citealt{maclow_klessen04}) that might partially explain the low level of star formation efficiency observed in the galaxies (e.g. \citealt{mckee07}).

GMCs are generally identified by contouring images above a certain column density, or flux levels when line emission data cubes are used. From observations, therefore, clouds are a set of connected pixels (either 2D or 3D) above a certain threshold level. These operations were done by eye in earlier studies (e.g. \citealt{dame86}). However, the use of \emph{position-position-velocity} (PPV) data cubes complicated the recognition of GMCs by eye. Therefore several automatic algorithms have been developed, able to handle the third dimension, as well as large datasets with different levels of blending between structures. Those algorithms are based either on iteratively fitting and subtracting a model to the molecular emission (as \verb"GAUSSCLUMPS", \citealt{stutzki_guesten90}; \citealt{kramer98}) or on the ``friends-of-friends'' paradigm that connects pixels according to their neighbors and values, without assuming a particular shape for the objects to decompose (as \verb"ClumpFind", \citealt{williams94} or \verb"CPROPS", \citealt{rl06}). Recently, gravity-based alternatives have also been proposed (\verb"Dendrograms", \citealt{rosolowsky08}; \verb"G-Virial", \citealt{li15}).  These latter approaches all assign individual pixels in a data cube to belong to single objects and GMC identification is thus a {\it segmentation} problem in the language of image processing.  Despite attempts to account for resolution and sensitivity biases (\citealt{rl06}), almost all algorithms for the cloud identification suffer from the influence of the survey design (e.g., \citealt{pineda09}; \citealt{wong11}). Depending on the complexity of the molecular environment, algorithms provide different results (\citealt{hughes13b}) and sometimes misleading ones (Schneider \& Brooks 2004). In particular, low resolution causes the blending of emission from unrelated clouds (as in M51, \citealt{colombo14a}); and high resolution makes segmentation algorithms identify cloud substructures as individual clouds.  In a clumpy medium, the friends-of-friends approach will naturally find objects with the scale of a few resolution elements.  The situation is further complicated in velocity-crowded regions like the Milky Way. 

In this paper, we consider the problem of GMC segmentation in the context of the more general theory of \emph{cluster analysis}. \emph{Clustering} is an unsupervised (no need for a training dataset) classification of patterns that groups sets of data in such a way that data in the same group (called a ``cluster'') are more similar to each other than to the data in other groups (``clusters''). Similarly, the process of finding GMCs in an image or in a data cube can be viewed as grouping pixels considered as part of a single entity as separated from others that are part of different entities. The concept of data clustering was originally introduced in anthropology by \cite{driver32}.  Clustering is now used by many disciplines to manage large quantities of data (data mining) or to reduce the data to learn information and make predictions (machine learning; for a general review about clustering, see \citealt{jain99}). Viewing GMC segmentation as a clustering problem allows us to create an algorithm able to overcome many of the limitations noted above (in particular the over-division of the CO emission caused by high resolution) and to generate physically-oriented cloud catalogs. Many clustering algorithms are based on graph theory (e.g., \citealt{jain99}). In Section~\ref{S:dendro} we show how graph representations of star-forming complexes are naturally provided by dendrograms. Dendrograms give a very detailed view of the global hierarchical structure within a molecular line data cube, but by themselves cannot be used to identify clouds. Nevertheless, graph abstraction furnishes a direct way to use the dendrogram for GMC segmentation.  We will introduce in Section~\ref{S:dendro} the graph theory basis for the problem and the algorithm chosen for the cluster analysis: \emph{spectral clustering} (Section~\ref{SS:spclust}). In Section~\ref{S:scimes}, we outline our method, \texttt{SCIMES} (Spectral Clustering for Interstellar Molecular Emission Segmentation) and the specific criteria that we use to extract discrete objects from the dendrograms of emission. In Section~\ref{S:test}, we show how the different segmentation criteria influence the final cloud decomposition using data from the Orion-Monoceros region. We demonstrate the stability of the method with respect to changing dendrogram parameters, noise realizations and dataset resolution. In Section~\ref{S:algo_comp}, we compare our method with other cloud decomposition algorithms, and we show how different segmentations produce different cloud properties in term of scaling relations and mass spectra. We examine how a cloud decomposition together with the knowledge of the hierarchical structure of the clouds might improve our understanding about the dynamical state of the clouds in the Orion-Monoceros complex (Section~\ref{S:scal_rel}). Finally we discuss the novel aspects and possibilities offered by the algorithm in Section~\ref{S:disc}. We summarize the paper content and results in Section~\ref{S:summary}.

%=====================================================================
\section{Using dendrograms to identify Giant Molecular Clouds}
\label{S:dendro} 
%=====================================================================
A dendrogram is a tree diagram that represents the hierarchy of structures within some data. A dendrogram is composed of two types of structures: branches, which are structures that split into multiple sub-structures, and leaves, which are structures that have no sub-structure. Branches can split up into branches and leaves, which allows hierarchical structures to be adequately represented. The term trunk is used to refer to a structure that has no parent structure. Dendrograms provide a precise representation of the topology of star-forming complexes. To use dendrograms to identify clouds, we need to interpret the dendrogram in the framework of graph theory and cluster analysis.

%---------------------------------------------------------------------
\subsection{Dendrogram in astronomy: definition and construction}
\label{SS:dendro_constr} 
%---------------------------------------------------------------------

In this paper we use the dendrogram implementation defined in Rosolowsky et al. (2008, hereafter: R08) that generalized the original concept of \cite{houlahanscalo92} to three dimensional data cubes including standard molecular line techniques to characterize the structures defined by the dendrogram itself. Here we give a brief description of the dendrogram technique that constitutes the core of the method we developed.

In astronomy, we define the \emph{dendrogram} or \emph{structure tree} as a ``stick man'' abstraction of the hierarchical structure of molecular gas (see figure~\ref{F:dendro_graph}; figure~\ref{F:dendro_weight} panels \emph{a}, and \emph{b}; and figure~\ref{F:orion_dendro}). It represents how the three dimensional contours (or isosurfaces at given emission levels) in a position-position-velocity molecular line data cube nest inside each other.

Following the terminology of \cite{houlahanscalo92}, a dendrogram is composed of \emph{leaves} and \emph{branches}. Leaves represent three-dimensional contours that contain a single local maximum and define the top of the dendrogram.  To suppress structure created by noise fluctuations, maxima are identified from all volumetric pixels in a data cube that have values larger than all of their neighbors over a box $D_{max}\times D_{max}\times\Delta V_{max}$, where $D_{max}$ and $\Delta V_{max}$ are set to some significant numbers of spatial and spectral resolution elements, respectively. The total number of identified local maxima is subsequently decimated to account for the effects of the noise, as follows. A local maximum is eliminated if the isosurface that contains it has a volume lower than a certain number of pixels ($N_{pix}$) and/or if its peak is lower than a certain brightness temperature difference, $\Delta T_{max}$, above the contour level where the local maximum merges with another local maximum. In this case, the emission profile that contains both local maxima is considered as a single object. Pairs of isosurfaces join at specific contour levels named \emph{merge} levels. A vertical line in the dendrogram that connects two leaves is called a ``branch''. The length of the branches represents the range of contour levels where the properties of the emission do not change significantly according to the connectivity criterion used (see R08 for more details). The implementation by R08 forces the dendrogram technique to generate only binary mergers, i.e. defined by the joining of two single objects. Eventually all branches and leaves in the dendrogram merge at a minimum temperature level to form the \emph{trunk} of the structure tree. The minimum temperature level is generally fixed to $n$ times the noise fluctuation level ($T_{min}=n\sigma_{\mathrm{rms}}$). 

%---------------------------------------------------------------------
\subsection{Interpreting the dendrogram as a graph}
\label{SS:dendro_graph} 
%---------------------------------------------------------------------
Although dendrograms are effective abstractions of the hierarchical structure of molecular emission, they cannot be used, by themselves, to identify molecular clouds. \emph{The main goal of this work is to provide a robust, mathematically-based method that finds the optimal cuts of a structure tree based solely on the properties of the data.}  The partitioned structure tree then defines discrete objects, if they exist, within data. To do so, we will study dendrograms in the broader framework of  graph theory, on which a large number of image analysis methods and clustering techniques rely.  We first introduce some basics of graph theory. 

\begin{figure}
{\includegraphics[width=0.4\textwidth]{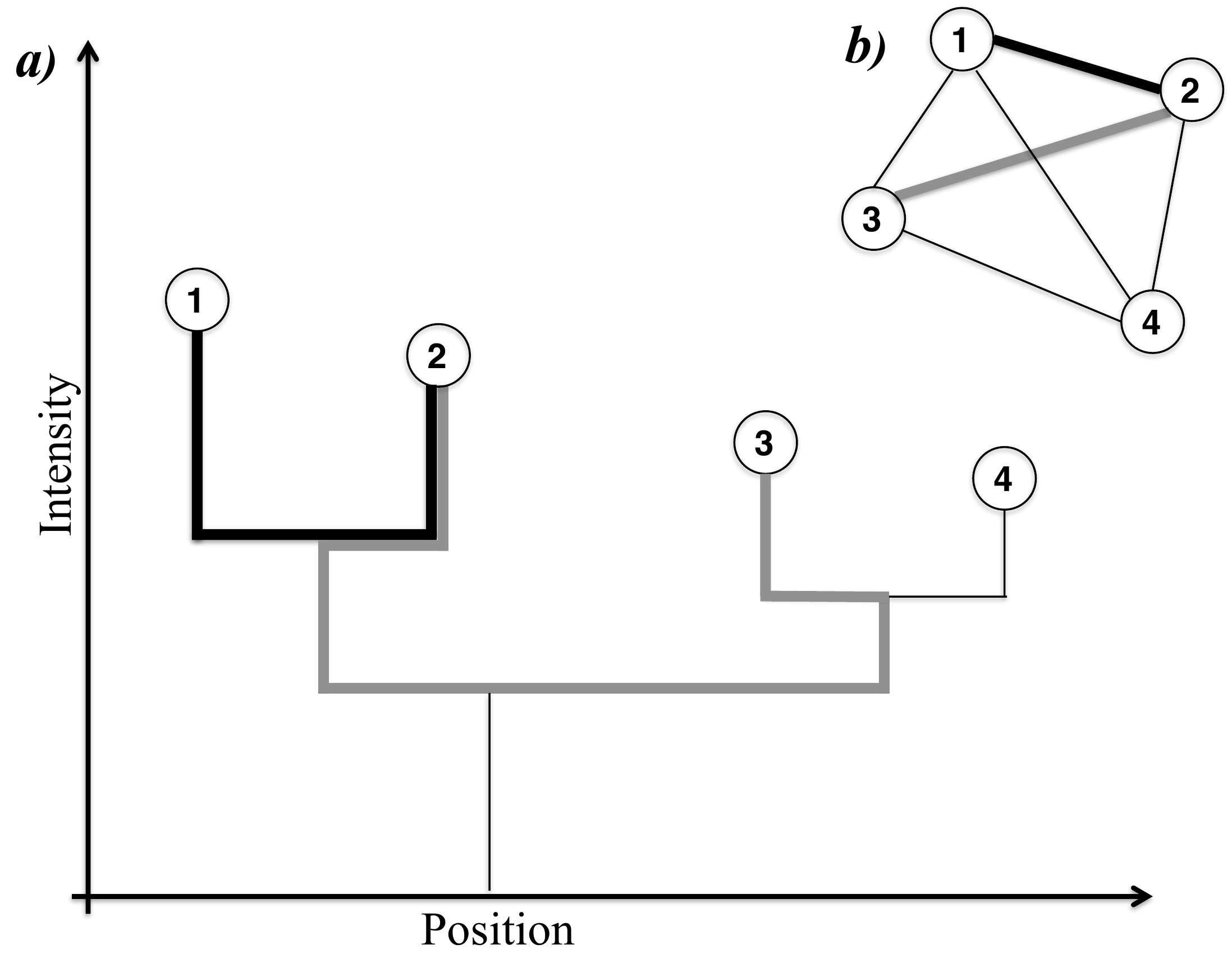}}
\caption{A dendrogram (main panel, \emph{a}) interpreted as a graph (inset panel, \emph{b}).  Leaves are seen as graph vertices, while edges are defined from the connection between pair of leaves. The weight of the edge between leaves connected at high hierarchical intensity level (e.g., leaves 1 and 2, black thick line) is higher than the weight at low hierarchical level (e.g, leaves 2 and 3, gray thick line).  The graph defined from a dendrogram is fully connected, since all vertices are, at minimum, related through the trunk.}
\label{F:dendro_graph}
\end{figure}

A graph is a mathematical entity defined as an ordered pair $G=(\mathcal{V},E)$ consisting of a set $\mathcal{V} = \{v_{1},..., v_{n}\}$ of ``vertices'' or ``nodes'' and a set $E$ of ``edges'', which are 2-element subsets of $\mathcal{V}$ (i.e., a single edge connects two vertices). Practically, vertices represent the group of objects we wish to cluster and edges represent the connections, links or the relations between those objects (figure~\ref{F:dendro_graph}\emph{b}).  

Dendrograms can be viewed as graphs by associating the leaves (local maxima) as the vertices whose edge is the largest-level isosurface containing both the leaves.   In the dendrogram, the highest branch where two leaves join is mapped to a graph edge.  Considering a pair of vertices $(v_{i},v_{j})$, dendrograms can be further described as  {\em weighted} graphs, where each edge has an associated non-negative value $w_{ij}\geq 0$, i.e., the connections between the objects have different ``strength''.  There are many possible choices for an edge weight, but in our application we use a weight given by the inverse of the merge level isosurface properties (we will clarify this aspect in Section~\ref{SS:spclust}).  A graph of a dendrogram for a single object is also {\em fully} or {\em strongly connected} where an edge exists between each pair of vertices, that is every vertex is connected to every other vertex in the graph or $w_{ij}>0$ always.  Thus, every pair of leaves is associated with a structure at a certain hierarchical level (see figure~\ref{F:dendro_graph}).  We consider the structures (leaves and branches) arising from the very bottom of the dendrogram all connected through an artificial ``super-structure'' that includes all of them: the trunk.  The isosurface associated with the trunk is given by the union of all the isosurfaces associated to the structures arising from the trunk\footnote{In R08, dendrogram branches without parental structures are called trunks. Here, we are interested in fully connected graphs, therefore we adopt always the name ``branch'' for structures that split into sub-structures independently by the parental structure and we consider a single trunk that contains all structure of the dendrogram.}.   Finally, dendrogram graphs are also {\em undirected}, i.e. the relations between vertices are symmetric ($w_{ij}=w_{ji}$).   In our application, we choose a symmetric weighting scheme since there is no apparent reason to consider a pair of leaves not connected on a one-to-one basis. Dendrogram graphs are also  {\em simple}, without self-loops ($w_{ii}=w_{jj}=0$) since we are interested only in the relations between pairs of leaves.

%=====================================================================
\subsection{Identifying objects in a dendrogram using spectral clustering}
\label{SS:spclust}
%=====================================================================
Having recast the dendrogram as a graph, we can identify objects within the dendrogram using one of the large class of graph-based clustering techniques.  Among these,  \emph{spectral clustering} works well on fully connected, weighted, undirected, and simple graphs such as those derived from the dendrogram.  Spectral clustering uses the eigenvectors of a matrix that parametrizes the relationship strengths (``similarity'') between the graph nodes to conduct dimensionality reduction before performing a standard clustering in fewer dimensions.  The clustering finds the optimal cut of the graph based upon the desired number of clusters ($k$), which must be provided as an input.\\  

\noindent The general algorithm of spectral clustering can be summarized in the following points (e.g. \citealt{vonluxburg07}):\\

\noindent \verb"Input": A \emph{similarity matrix} $S$ such that $s_{ij}$ is the similarity between the $i$th and the $j$th vertex , and $k$ is the number of clusters to generate.  For the GMC segmentation problem, we describe how to construct $S$ and choose $k$ in sections \ref{SS:scimes_gmcaff} and \ref{SS:scimes_sigma}, respectively.

\begin{figure*}
{\includegraphics[width=0.5\textwidth]{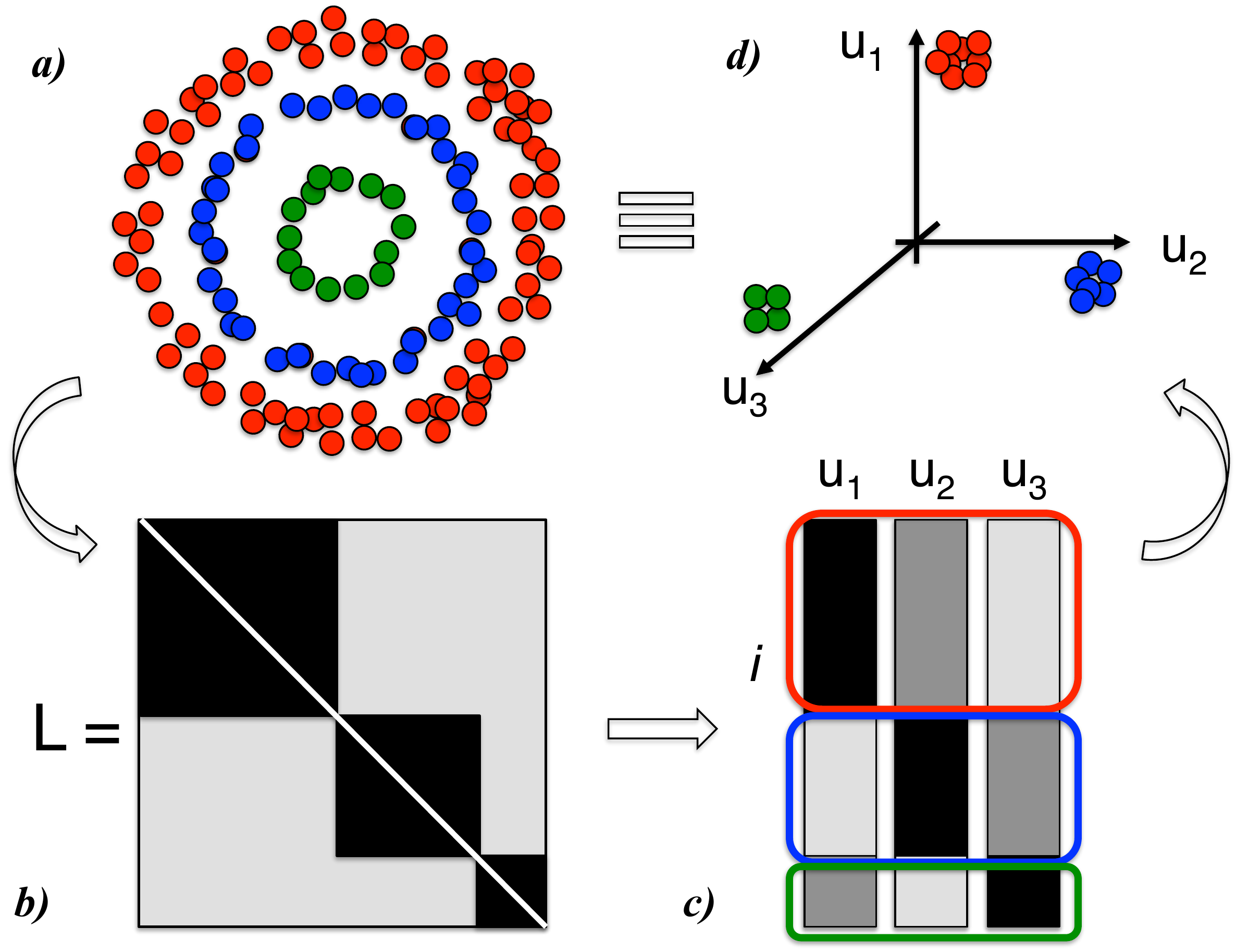}}
\caption{Graphical representation of spectral embedding. \emph{a)} Concentric distributions of objects. Different color indicates which groups of objects have high similarity. \emph{b)} Graph Laplacian for objects in the previous step. Matrix elements that represent pairs of objects with high similarity are drawn in black, while low similarity between objects is indicated in grey. The main diagonal of the graph Laplacian contains the degrees of the graph vertices (equation \ref{E:vdeg}). When an appropriate similarity criterion is chosen for clustering, the Laplacian matrix appears block diagonal (after an index permutation). \emph{c)} The eigenvectors obtained after the spectral embedding, chosen from the largest $k$ eigenvalues.   The number of eigenvectors considered defines the dimension of the clustering space ($\mathbb{R}^{k}$) and is equivalent to the number of the desired clusters.  In this case the number of clusters $k=3$ and every object (or vertex) of the initial graph is represented as a point in $\mathbb{R}^{3}$ with coordinates [$u_{1}(i),u_{2}(i),u_{3}(i)$]. \emph{d)} Embedded clustering space. The initial distributions of objects are well separated in this space, enhancing the similarities between objects. Objects in this space can be easily clustered using \emph{k-}means and Euclidean distances.}
\label{F:spclust}
\end{figure*}

\begin{enumerate}

\item Construct the \emph{degree matrix} $D$ and the \emph{graph Laplacian} $\mathcal{L}$ (section \ref{SSS:spclust_simmat}, see also figure~\ref{F:spclust}b).

\item Compute the \emph{spectral embedding}, i.e. calculate the first $k$ larger eigenvalue eigenvectors $u_{1},..., u_{k}$ of $\mathcal{L}$ (section \ref{SSS:spclust_embed}, see also figure~\ref{F:spclust}c).

\item Construct the matrix $U\in\mathbb{R}^{n\times k}$ made by the $k$ eigenvectors $u_{1},..., u_{k}$ as columns (section \ref{SSS:spclust_embed}, see also figure~\ref{F:spclust}c).

\item Let $y_{i}$ be points in $\mathbb{R}^{k}$ where $i=1,...,n$, corresponding to the $i-$th row of $U$ (section \ref{SSS:spclust_embed}).

\item Cluster the points $(y_{i})_{i=1,...,n}$ in $\mathbb{R}^{k}$ with the \emph{k-means} algorithm into clusters $(C_{\ell})_{\ell=1\dots k}$  (section \ref{SSS:spclust_kmeans}, and figure~\ref{F:spclust}d).

\end{enumerate}

\noindent \verb"Output": Clusters $A_{1},...,A_{k}$, which are sets of vertices in the original space such that $v_i\in A_{\ell}$ if $y_{i}\in C_{\ell}$.\\

\noindent Popular variations of this algorithm can be found in \cite{shi_malik00} and \cite{ng01}. In the following we will explain each step the spectral clustering algorithm, occasionally sacrificing the mathematical formalism in favor of intuition.

%---------------------------------------------------------------------
\subsubsection{Graph representations as matrices: graph Laplacian and degree matrices}
\label{SSS:spclust_simmat} 
%---------------------------------------------------------------------
The strength of the relation present between two nodes of the graph can be seen as the \emph{similarity} that exists between them. In the most general sense this concept can also be related to the notion of distance: higher ``similarities'' between two objects imply that the ``distance'' between them is shorter. In graph theory, a similarity between a pair of vertices is quantified by the weight of the edge that connects the vertices. All the similarity between each pair of nodes in the graph can be collected in a \emph{similarity matrix} that further abstracts the graph and constitutes the main input of spectral clustering algorithms (see figure~\ref{F:dendro_weight}c and figure~\ref{F:orion_affmats}). The \emph{similarity matrix} (also called \emph{affinity} or \emph{adjacency} matrix), therefore, parametrizes the quantitative assessment of the relative similarity of each pair of vertices in the graph. For convenience we introduce the shorthand notation $i\in S$ for the set of indexes $\{i|v_{i}\in S\}$, where $S=(s_{ij})_{i,j=1,...,n}$ represents the affinity matrix we are dealing with and $n$ is the number of objects or graph vertices. The affinity matrix provides a natural representations of the graph; therefore in the case of a dendrogram-derived graph, $S$ is square ($S\in\mathbb{R}^{n\times n}$), symmetric ($s_{ij}=s_{ji}$), with null main diagonal ($s_{ii}=0$; graph simplicity requirement) and positive semidefinite (graph strong connectivity requirement). Since the graph represents the local neighborhood relationships, the affinity matrix itself should reflect the local neighborhoods. 

To accomplish this, the affinity matrix is usually rescaled using a \emph{kernel} function\footnote{In clustering analysis literature this operation is called ``smoothing''. Nevertheless, in the text we use expression ``rescaling'' to indicate the affinity matrix smoothing, in order to avoid confusion with the image smoothing concept generally used in astronomy.}.  A Gaussian kernel is commonly used: 

\begin{equation}\label{E:gauss_ker}
s_{ij}=\exp\left(\frac{-w_{ij}^{2}}{2\sigma_{S}^{2}}\right), 
\end{equation}

\noindent where the parameter $\sigma_{S}$ controls the size of the neighborhoods and must be carefully chosen\footnote{\texttt{SCIMES} uses a modified version of the Gaussian kernel proposed by \cite{shi_malik00}, i.e. $s_{ij}=\exp(-w_{ij}^{2}/\sigma_{S}^{2})$ that produce more restrictive rescaling of the affinity matrix.}.  Identifying an appropriate affinity matrix represents the most challenging task of the spectral clustering technique. Affinity matrices can be potentially constructed using almost any property that can be seen as similarity or distance. The choice of similarity criterion together with the choice of $\sigma_{S}$ influence the quality of the clustering partition we obtain. As an heuristic, good affinity matrices appear ``block diagonal'' (after applying appropriate row/column permutations) where the values on the boundary of each block is similar.

Most of spectral clustering-based algorithms make use of a different ``form'' of the affinity matrix called graph Laplacian (\emph{(i)}-paragraph of the spectral clustering general algorithm in Section \ref{SS:spclust}), since its properties are more suitable for the spectral embedding.
The unnormalized form of the graph Laplacian $\mathcal{L}$ is defined as $\mathcal{L}\equiv D-S$, where $D$, called \emph{degree matrix}, is a diagonal matrix that contains the degrees $d_{i}$ of the vertices $v_{i}$ on the main diagonal. The degree of a vertex $v_{i}$ is defined as:

\begin{equation}
\label{E:vdeg}
d_{i} \equiv \sum_{j=1}^{n}s_{ij}. 
\end{equation}

 Often, the ``symmetric normalized'' form of the Laplacian is used: $\mathcal{L}_{\mathrm{sym}} \equiv D^{-1/2}(D-S)D^{-1/2}$ (e.g., \citealt{ng01}), since it produces more general eigenvalue, better related to other graph invariants, and with a direct connection to spectral geometry and in stochastic processes (\citealt{chung97}).

\noindent The graph Laplacian fully represents the algebraic properties of the graph.   The utility of the graph Laplacian can be understood by considering a simpler type of graph that is unweighted and weakly connected (i.e., there are disconnected parts of the graph).  The similarity matrix is then binary where $s_{ij}=1$ if there is an edge between $v_i$ and $v_j$ and $s_{ij}=0$ otherwise.  Then, $d_i$ is just the number of nodes connected to $v_i$.  The Laplacian then has the degree along the diagonal and $l_{ij}=-1$ indicating a connection between $v_i$ and $v_j$.  In this view, the graph Laplacian is the discrete version of the continuous Laplacian operator $\nabla^2$ (i.e. the multi-variable second-derivative), operating on the graph.  Denser nodes are equivalent to ``bumps'' in the second derivative of a continuous function.  Several spectral features of the graph Laplacian are very useful to quickly assess the global properties of the graph it represents. For example, the number of 0-valued eigenvalues of $\mathcal{L}$ corresponds to the number of graph's connected components (i.e. groups of nodes or ``clusters''). Indeed, each connected component forms a ``block'' in the Laplacian matrix (after appropriate permutations), therefore, the nodes of these components only have edges within themselves. Each of these groups can be represented by a fully connected graph and their graph Laplacian has only a single eigenvalue equal to 0.  Since the graph Laplacian is also positive-semidefinite, the second smallest eigenvalue is greater than zero.  This eigenvalue is the {\em algebraic connectivity} of the graph and quantifies how well the graph is connected.    

%---------------------------------------------------------------------
\subsubsection{Spectral embedding}
\label{SSS:spclust_embed} 
%---------------------------------------------------------------------

The main utility of the spectral clustering is to map the data represented as a graph to a different vector space where the cluster properties of the data (if they exist) are enhanced.  This is accomplished thanks to the properties of the graph Laplacian through the \emph{spectral embedding} (second and third points of the spectral clustering general algorithm in Section \ref{SS:spclust}) that changes the representation of the data points to points $y_{i}\in\mathbb{R}^{k}$. The elements of first $k$ eigenvectors provide a lower-dimensional description of the block diagonal structure of the Laplacian (or the similarity matrix) and of the $k$ connected components of the graph. A graphical description of this concept is reported in figure \ref{F:spclust}.
%while the mathematical theorems and proofs can be found in Appendix (NOT YET, BUT NEEDED AT ALL?). 

%---------------------------------------------------------------------
\subsubsection{\emph{k}-means algorithm}
\label{SSS:spclust_kmeans} 
%---------------------------------------------------------------------

The data to cluster are mapped through the spectral embedding into $y_{i}$ points of $\mathbb{R}^{k}$. In this new ``clustering'' space, abstract description of similarity between vertices are translated into Euclidean distance. The data in this space are easily clustered with common clustering algorithms such as \emph{k}-means that find groups where the intra-cluster distance is maximized while the inter-cluster similarity is minimized, given the desired number of clusters $k$. The \emph{k}-means algorithm (\citealt{macqueen67}) is the most popular algorithm for clustering given its conceptually simple idea and its fast convergence speed. The algorithm works in $\mathbb{R}^{n}$ randomly or using some heuristic information (\citealt{arthur07}) by selecting $k$ ``means'' or ``centroids'' of the data to cluster, where the number of clusters $k$ is provided as input (see Section~\ref{SSS:spclust_clnum} for details\footnote{In this work we use a heuristic version of k-means, \emph{k-means}$++$ by \cite{arthur07} that optimizes the initial seeding of the random centers.}). Generally, $k\ll n$, where $n$ is the number of objects to cluster, in this case the graph vertices or the dendrogram leaves. The $k$ clusters are then generated by associating every observation with the nearest mean. The centroids of each clusters are subsequently used as new ``means''. The last two steps are iterated until convergence. The convergence is reached once the current centroids are at the same positions of the previous ones. The choice of the \emph{k}-means algorithm for identifying clusters is not fundamental.  If the similarity function is appropriate for clustering the data, any cluster algorithm can be implied to obtain the final product.

\begin{figure}
\begin{center}
\includegraphics[width=0.4\textwidth]{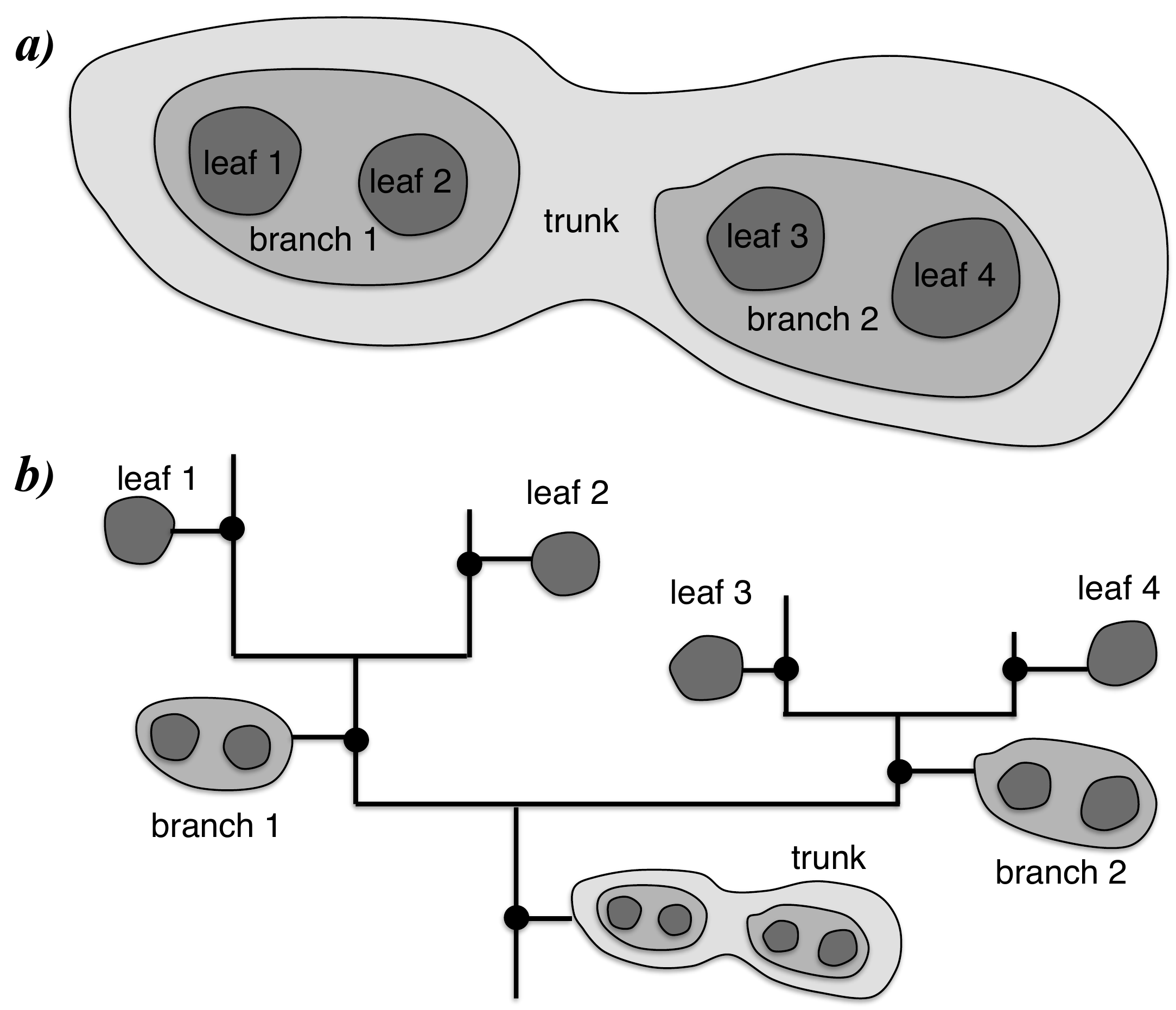}
\includegraphics[width=0.4\textwidth]{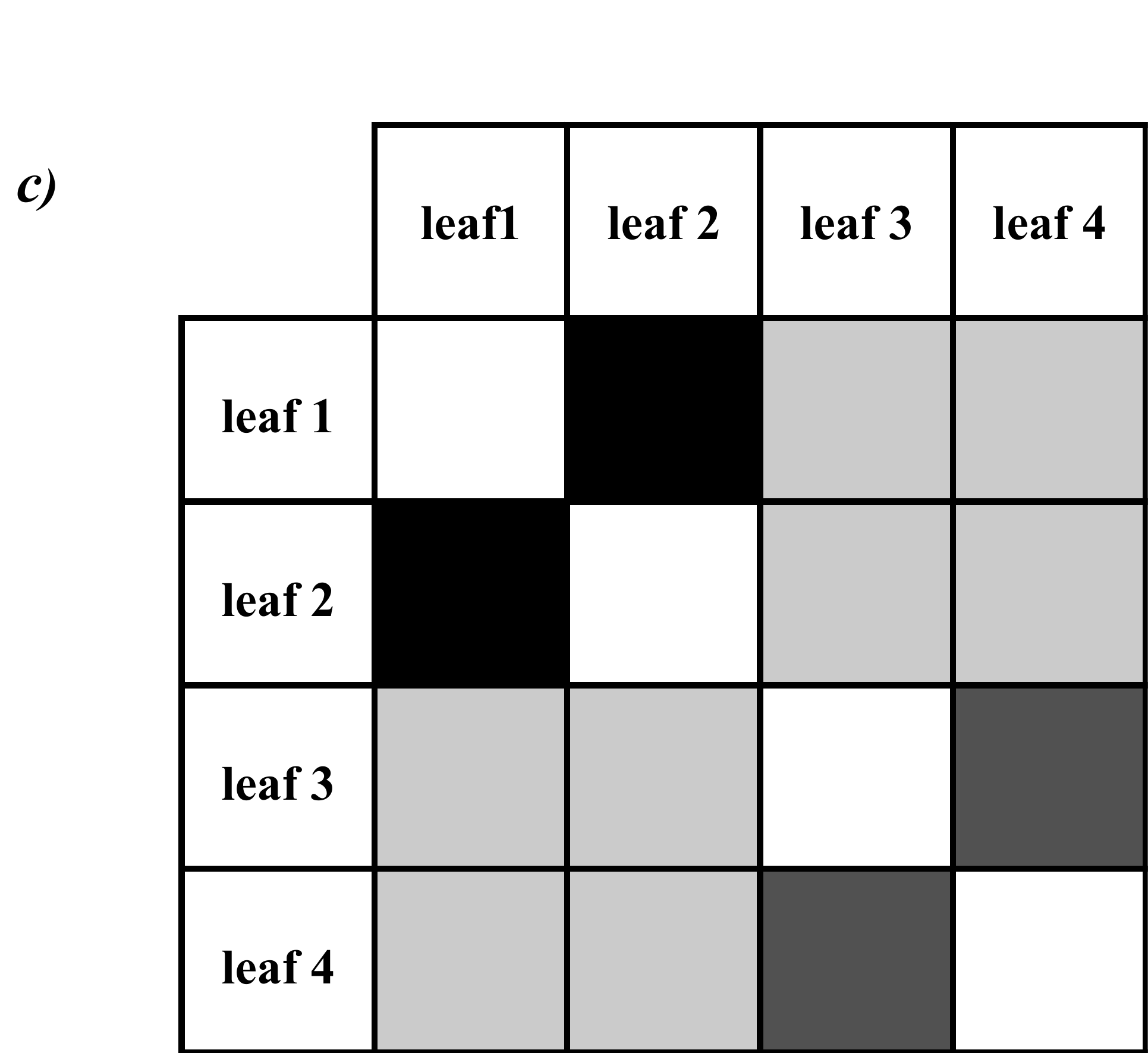}
\end{center}
\caption{From the molecular line emission feature \emph{(a)} to the dendrogram and \emph{(b)} the similarity matrix \emph{(c)}. Leaves 1 and 2 join at branch 1, as well as leaves 3 and 4 join at branch 2. However leaves 1 and 2 are also connected to leaves 3 and 4 through the isosurface correspondent to the trunk at lower hierarchical level. The weight of the edges between each pair of leaves is defined as the inverse of the properties of isosurface embedded molecular emission where to two leaves ``join''. In this example we consider the ``area'' of the isosurface to weight the edges: the weight of the edge between leaves 1 and 2 is similar to the weight of the edge between 3 and 4, however the edge weights between leaves 1 and 3, 1 and 4, 2 and 3, 2 and 4 is much lower, since the embedding area is much larger. The similarity matrix \emph{(c)} contains those weights. In the picture, darker colors indicate higher similarity. According to this matrix, an optimal graph partition for 2 clusters might provide the two objects identifiable with branches 1 and 2.}
\label{F:dendro_weight}
\end{figure}

%---------------------------------------------------------------------
\subsubsection{Evaluate the number of clusters}
\label{SSS:spclust_clnum} 
%---------------------------------------------------------------------
A common problem for several clustering algorithms is to select the number of clusters $k$ to be generated, which must be provided as an input.  Given this number, the algorithms proceed to find the best arrangement of the data within $k$ groups. Various methods have been proposed to estimate $k$ ranging from theoretical approaches (\citealt{still_bialek04}), to gap statistics (\citealt{tibshirani00}), and stability approaches.  For spectral clustering, the number of clusters can be guessed by analyzing the eigenvalues and the properties of the eigenvectors themselves (e.g. \citealt{zelnik_manor04}). Other methods aim to assess the quality of the clustering using measurements of the ratio of within-cluster and between-cluster similarities. An example of such a measure is the \emph{silhouette} (\citealt{rousseeuw87}). The silhouette coefficient is defined as:

\begin{equation}\label{E:silhouette}
\mathrm{sil}(i) = \frac{b(i)-a(i)}{\max(a(i),b(i))},
\end{equation} 

\noindent where $a(i)$ represents the average similarity between the object $i$ and all other elements in the same cluster, $b(i)$ is the average similarity between $i$ and all other elements in the next nearest cluster and $-1\leq \mathrm{sil}(i)\leq 1$. Therefore the silhouette directly relates with the general definition of ``clustering'' (see section \ref{S:intro}): in particular, $\mathrm{sil}(i)=1$ for dense (high intra-cluster similarity) and well separated clusters (low inter-cluster similarity), $\mathrm{sil}(i)=-1$ for incorrect clustering and $\mathrm{sil}(i)=0$ for overlapping clusters\footnote{Spectral clustering does not allow $\mathrm{sil}(i)=0$ since it uses a hard assignment method: an object can not belong to two different clusters.}.  The average $\mathrm{sil}(i)$ over all data of all clusters is a  measure of how well the data have been partitioned and how appropriately $k$ has been chosen. The average silhouette is not a monotonic function of $k$ so the best number of clusters $k$ is determined by maximizing the silhouette.

%=====================================================================
\section{SCIMES - Spectral Clustering for Interstellar Molecular Emission Segmentation}
\label{S:scimes} 
%=====================================================================
Having introduced the mathematical framework needed to convert dendrograms into graphs and to optimally cut these graphs through spectral clustering, we now present our algorithm, \texttt{SCIMES}, to identify significant objects within the structure tree of molecular emission. \texttt{SCIMES} uses the dendrogram as input produces different properties (as effective radius, velocity dispersion, flux etc.) associated with the structures within the dendrogram. We observed that the \emph{luminosities} of the emission within the isosurfaces and \emph{volumes} of the isosurfaces are good criteria to define the similarity matrices at the top hierarchy between each pair of leaves (Section~\ref{SSS:scimes_fluxvol}). The affinity matrices are rescaled using a kernel through an appropriate choice of $\sigma_{S}$ (Section~\ref{SS:scimes_sigma}). The rescaled similarity matrices can be aggregated to obtain a cluster configuration that depends on all chosen affinity criteria. Then an approximate number of clusters, $k_{g}$, is estimated through a direct analysis of the final affinity matrix (Section~\ref{SSS:scimes_mataggr}).  The degree matrix and Laplacian are automatically defined as described in Section~\ref{SSS:spclust_simmat}. Nevertheless, the best number of clusters $k_{b}$ is ultimately defined through the silhouette (see Section~\ref{SSS:spclust_clnum}), running the spectral clustering algorithm several times, such that $k_{g}-15\leq k_{b}\leq k_{g}+15$. Finally, the clusters that do not correspond to single branches in the dendrogram are pruned and the remaining clusters are labeled to obtain the GMCs (Section~\ref{SS:scimes_rem}).\\ 

\noindent The dendrogram and the catalog of the structures within it (\texttt{SCIMES} inputs) are defined from a molecular line data cube using the python distribution \verb"ASTRODENDRO" (http://www.dendrograms.org/). This dendrogram implementation package requires setting three input parameters: \verb"min_value" below which any value is not considered in the dendrogram construction (usually set to several times the sensitivity level of the dataset $\sigma_{\mathrm{rms}}$); \verb"min_delta" indicating how significant a leaf must be to be considered independent (again set equal to several times the observation sensitivity); \verb"min_pix", the minimum number of pixels needed for a leaf to be independent (generally equal to the several times the observation beam). We use the spectral clustering and silhouette implementations by \verb"SCIKIT-LEARN" (http://scikit-learn.org/stable/modules/clustering). In the following Sections, we describe the different steps of the algorithm, connecting back to the formalisms summarized in Section~\ref{S:dendro}.

%---------------------------------------------------------------------
\subsection{Similarity criteria for Giant Molecular Cloud segmentation}
\label{SS:scimes_gmcaff} 
%---------------------------------------------------------------------

Defining good similarity criteria is the most important step of the clustering process, since the algorithm finds the optimal graph cuts based solely on the ``features'' of the similarity matrix\footnote{The number of clusters $k$, the secondary input of the spectral clustering algorithm, is also automatically guessed by SCIMES from the similarity matrix.} and does not provide, at priori, any metric to understand the quality of the final clusters. 

By default \texttt{SCIMES} constructs the affinity matrices used for the clustering based on the ``volume'' and/or ``luminosity'' of the structures identified by the dendrogram. In the following, we describe the definition of these measurements and how to generate their associated similarity matrices.

%---------------------------------------------------------------------
\subsubsection{The luminosity and volume criteria}
\label{SSS:scimes_fluxvol} 
%---------------------------------------------------------------------
We define the edges of the dendrogram-derived graph from the properties of the largest-valued isosurfaces containing pairs of local maxima (the vertices; see also Section \ref{SS:dendro_graph}). Those isosurfaces contain molecular emission, and the properties of that emission are used to weight the edges of the graph.  To calculate these properties, the dendrogram implementation by R08 assumes the moment method \citep{rl06}. In this view, the $i$th pixel in a data cube can be identified with a brightness temperature $T_{i}$, positions $x_{i}$, $y_{i}$, $v_{i}$ and sizes $\delta x$, $\delta y$, $\delta v$ for the two spatial dimensions and the velocity dimension, respectively. Therefore, the \emph{flux} of the region corresponds to the zeroth moment, or the sum of all emission within the isosurface:

\begin{equation}
\label{E:flux}
F = \sum_{i} T_{i}\delta x\delta y\delta v.
\end{equation}

\noindent The flux can be converted into luminosity assuming a physical distance $d$ (in parsecs) to the target, yielding $L=Fd^{2}$, the first clustering criterion used by our code.
 
\noindent An isosurface has several morphological properties. To evaluate these properties, the major axis of the spatial projected structure is first located using the principal component analysis. The spatial axes are rotated such that the major axis of the region is aligned with the $x$ axis, while the minor axis with the $y$ axis. The root-mean-squared (rms) sizes of the region are estimated from the intensity-weighted second moments along the two spatial dimensions:

\begin{equation}
\label{E:majmin}
\sigma_{\mathrm{maj}}=\sqrt{\frac{\sum_{i}T_{i}(x_{i}-\overline{x})^{2}}{\sum_{i}T_{i}}}, \sigma_{\mathrm{min}}=\sqrt{\frac{\sum_{i}T_{i}(y_{i}-\overline{y})^{2}}{\sum_{i}T_{i}}};
\end{equation}

\noindent where the sum runs over all pixels within the isosurface. Combining the two measurements, the rms size is then:

\begin{equation}
\label{E:sigmar}
\sigma_{r} = \sqrt{\sigma_{\mathrm{maj}}\sigma_{\mathrm{min}}}.
\end{equation}

\noindent The radius of the spherical cloud can be related to $\sigma_{r}$ through $R=\eta\sigma_{r}$ where $\eta=1.91$ (\citealt{solomon87}, \citealt{rl06}). 

The velocity dispersion is calculated as:
\begin{equation}
\label{E:vrms}
\sigma_{v}=\sqrt{\frac{\sum_{i}T_{i}(v_{i}-\overline{v})^{2}}{\sum_{i}T_{i}}}.
\end{equation}
\noindent We use as second similarity criterion the \emph{volume} of the isosurface in PPV space:
\begin{equation}
\label{E:volume}
V = \pi R^{2}\sigma_{v}.
\end{equation}

Luminosity and volume have been chosen as the default clustering criteria for \texttt{SCIMES} for several reasons. First, those criteria are directly related to the basic physical descriptors of the molecular emission structures (morphology, velocity, and emissivity) and allow us to consider the structure neighborhood in both spatial and spectral directions, and in terms of CO emission differences. Second, luminosity and volume are monotonic and discontinuous properties of the isosurface related to the dendrogram. Those values increase monotonically against a decrease in the dendrogram hierarchy
level and produce large ``jumps'' in the affinity when two objects with similar volume of luminosity merge at a certain level (figure~\ref{F:orion_affmats}). By construction, the dendrogram is a monotonic structure, so the number of isolated isosurfaces should increase with the hierarchical level and it is easily recast into affinity matrices through the luminosity and the volume of its structures. Those features of luminosity and volume criteria give well-behaved block diagonal similarity matrices that are preferred when working with spectral clustering (see next Section). The scaling parameter of the rescaling
kernel (described in Section~\ref{SSS:spclust_simmat}) can be directly calculated from those kind of matrices; the number of clusters is easily guessed and, in general, corresponds to the number of blocks (Section~\ref{SS:scimes_sigma}). 

\texttt{SCIMES} accepts as input any kind of user-defined affinity matrix. Nevertheless, the code might not behave as expected when non-monotonic and strictly continuous criteria are used. In this aspect, volume and luminosity can be associated to the number of volumetric pixel within a certain isosurface, and to the sum of the values of them, respectively. However, those properties are largely continuous functions of the dendrogram hierarchy level. This makes the estimation of scaling parameter and initial number of clusters difficult, providing unwanted mergers between the structures.

Luminosity and volume criteria are general, since they embody, by definition, distance information. However, for several applications, especially involving data of the Galaxy, distances are rarely known. This entails some changes in the cloud segmentation provided by \texttt{SCIMES}. We discuss this issue in Appendix~\ref{A:nodist}.

\begin{figure}
{\includegraphics[width=0.5\textwidth]{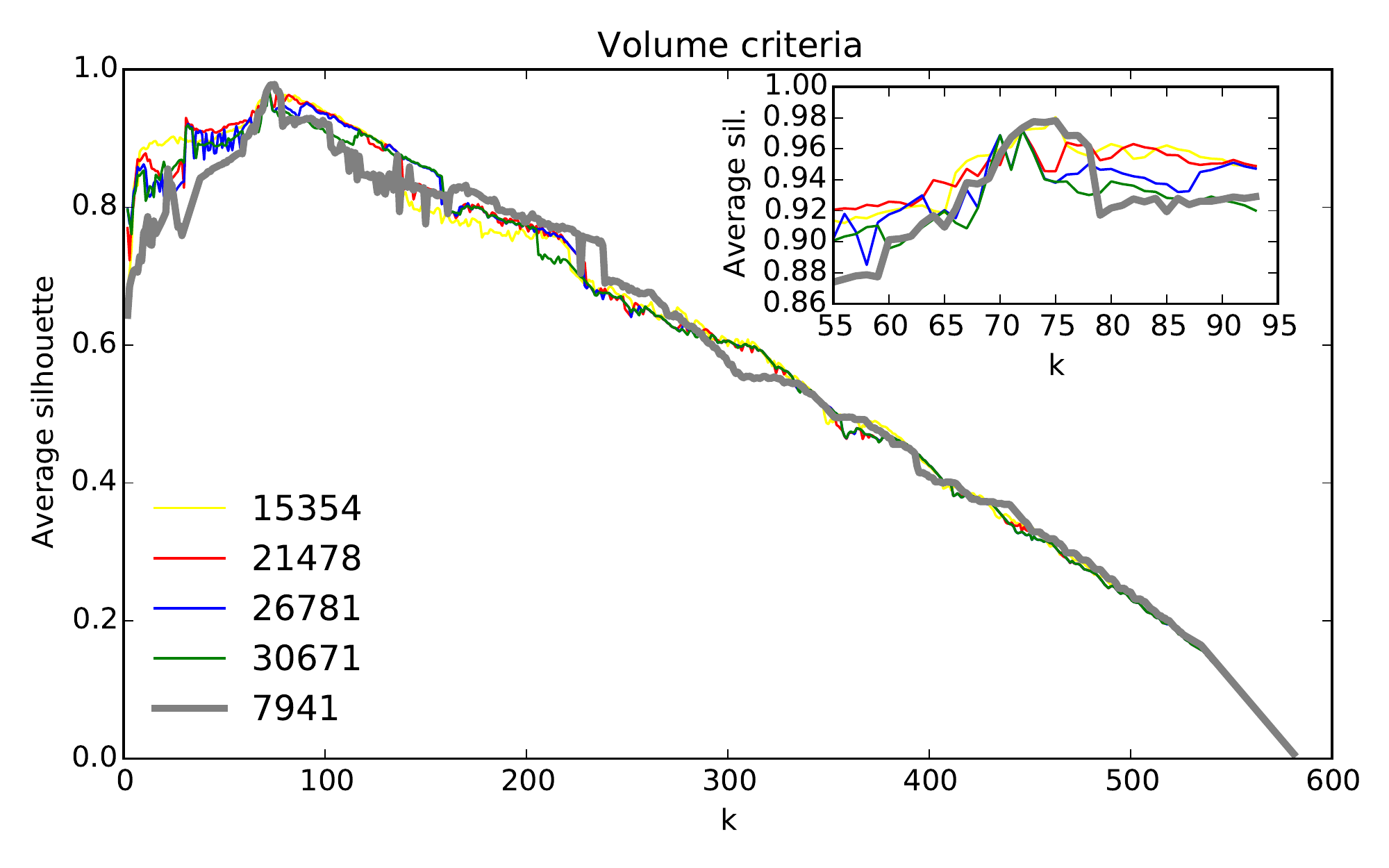}}
{\includegraphics[width=0.5\textwidth]{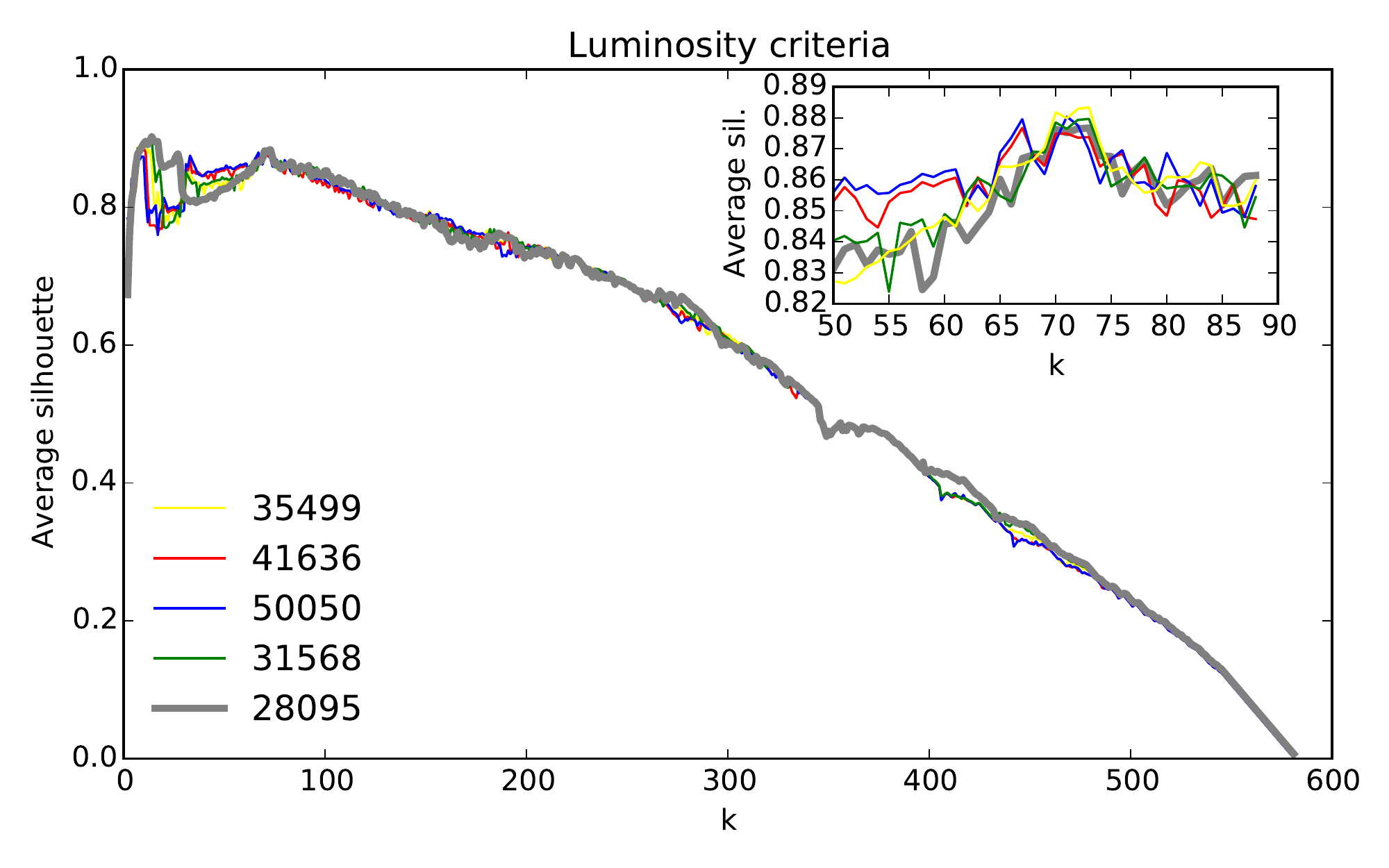}}
\caption{Average silhouette versus the number of clusters $k$ to determine the best number of clusters $k_{b}$, for selected values of $\sigma_{S}$. The behavior of the average silhouette is quite similar for both volume and luminosity criteria, although the profile for the luminosity criterion presents some ambiguities. Nevertheless, the dependency on the value of $\sigma_{S}$ is not as strong as expected. The values of $\sigma_{S}$ chosen using the \emph{similarity histogram} (see Section~\ref{SS:scimes_sigma} and figure~\ref{F:orion_amshisto} for details) are indicated in the bottom left corner of each panel. The values are in pc$^{2}$\,km\,s$^{-1}$ for the volume criteria (upper panel) and K\,km\,s$^{-1}$\,pc$^{2}$ for the luminosity criteria (lower panel). The profile of the chosen scaling parameters for the segmentation of the Orion-Monoceros dataset is indicated with gray thick lines. The insets at the top right corners of the plots show zoomed versions of the average silhouette profile around the peak.}
\label{F:orion_siltest}
\end{figure}

%---------------------------------------------------------------------
\subsubsection{From the similarity criteria to the similarity matrices}
\label{SS:scimes_smoaggr} 
%---------------------------------------------------------------------
Having identified the similarity criteria we found useful to partition the molecular line emission, here we explain how to construct the related similarity matrices. We already showed that dendrograms can be seen as fully connected, simple, undirected and weighted graphs in Section \ref{SS:dendro_graph}. The weight of the edges is determined by the properties of the highest hierarchical level in term of brightness temperature at which a pair of leaves (graph vertices) ``join'' (see figure \ref{F:dendro_graph}). Each merging level corresponds to molecular emission bounded by an isosurface. Along the branches the properties do not change much and are continuous functions of the contour level.  However where two branches merge, the properties change suddenly since the merged object contains more emission (see figure \ref{F:dendro_weight}). In general, higher hierarchical levels correspond to smaller isosurfaces and vice versa. Therefore the weight of the edges will be equivalent to the inverse of the properties of the molecular emission embedded by the isosurface at that particular hierarchical level. Considering two graph vertices labeled as $i,j$ (i.e., a pair of leaves in the dendrogram), we define the weight of the edge between them as 

\begin{equation}\label{E:weight}
w_{ij}=1/p_{ij},
\end{equation} 

\noindent where $p_{ij}$ indicates a property of the emission bounded by the highest-level isosurface containing the vertices. For the similarity criteria we established $p_{ij}$ as either $p_{ij}=L_{ij}$ or $p_{ij}=V_{ij}$.\\

%---------------------------------------------------------------------
\subsubsection{Similarity matrix aggregation}
\label{SSS:scimes_mataggr} 
%---------------------------------------------------------------------

\noindent Once all matrices have been rescaled using the kernel with the appropriate $\sigma_{S}$ (see Section~\ref{SS:scimes_sigma} for details), the matrices can be also \emph{aggregate} into a single similarity matrix that embodies all the similarity criteria. This process follows the idea of \cite{shi_malik00} for image segmentation.  They construct two similarity matrices for their problem (color image segmentation). After the rescaling the two matrices are multiplied element-wise. In the same way, we multiply our (volume and luminosity) kernel-rescaled similarity matrices element by element. The volume, luminosity, and/or the unique aggregate similarity matrices constitute the main input for the spectral clustering algorithm.

\begin{figure}
{\includegraphics[width=0.5\textwidth]{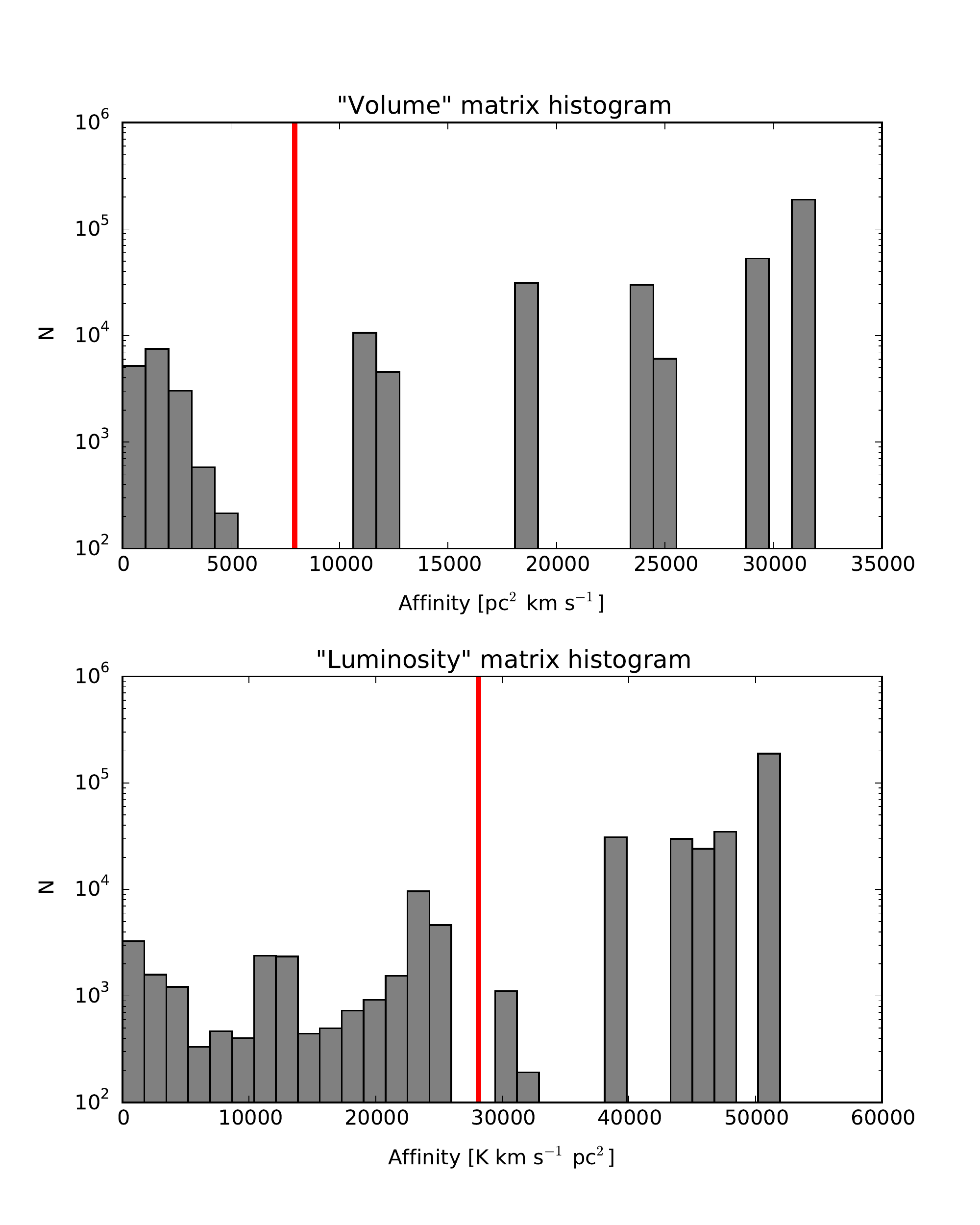}}
\caption{The multi-modal ``similarity histogram'' obtained from the volume (upper panel) and luminosity (lower panel) affinity matrices. $N$ indicates the number of affinities with a specific value. The values of $\sigma_{S}$ used for the average silhouette test in figure~\ref{F:orion_siltest} and section~\ref{SS:scimes_sigma} correspond to the average value between the different modes in the histograms. The chosen scaling parameters for the segmentation of the Orion-Monoceros dataset are indicated with red lines.} 
\label{F:orion_amshisto}
\end{figure}

%---------------------------------------------------------------------
\subsection{Guessing the scaling parameter and the number of clusters}
\label{SS:scimes_sigma} 
%---------------------------------------------------------------------

\noindent As for the choice of the right affinity criteria, selecting an optimal scaling parameter $\sigma_{S}$ is essential because it might significantly affect the number of the clusters identified and the quality of the clustering. Indeed, the scaling parameter determines how rapidly the similarity $p_{ij}$ falls off with the distance between leaves $i$ and $j$. Given the assumed rescaling kernel (Section \ref{SSS:spclust_simmat}), a large $\sigma_{S}$ merges the clusters resulting in an undesirable clustering; but a too-small $\sigma_{S}$ generates a weak similarity matrix where only the affinities of directly neighboring leaves are high.  The graph theory literature does not provide firm criteria to select good scaling parameters. \cite{ng01} suggest that the right $\sigma_{S}$ can be determined by evaluating the tightness of the clusters on the surface of a sphere.  This criterion deals with the quality of the clustering, therefore the ``tightness'' of the clusters on the surface of a sphere can be determined, similar to how the number of clusters $k$ is set using the \emph{silhouette} method (Section~\ref{SSS:spclust_clnum}).   To test this method, we ran the spectral clustering algorithm on the Orion-Monoceros dataset (see Section~\ref{S:test}) several times with different values of scaling parameters (figure~\ref{F:orion_amshisto}); and for all possible number of clustering configurations, i.e. $2\leq k\leq n-1$, where $n$ is the number of leaves. The best clustering configuration is given by $(k_b,\sigma_{S,b}) = \mathrm{argmax} [\mathrm{sil}(k,\sigma)]$.  Figure~\ref{F:orion_siltest} shows the results of the test. The average silhouette for both criteria presents a similar behavior. Nevertheless, the index has a clear peak for the volume criteria at $k=76$, while there is some ambiguity regarding the luminosity criteria, since the average silhouette profile presents two peaks at $k=14$ and $k=69$. While selecting $k=69$ for the luminosity criterion produces a clustering configuration similar to the volume criterion, the choice $k=14$ merges Orion A, NGC2149 and Monoceros into a single object. Surprisingly we note that the value of the silhouette, in general is not highly influenced by the selection of $\sigma_{S}$, but varies significantly according to the selected similarity criteria. 

\begin{figure*}
{\includegraphics[width=0.75\textwidth]{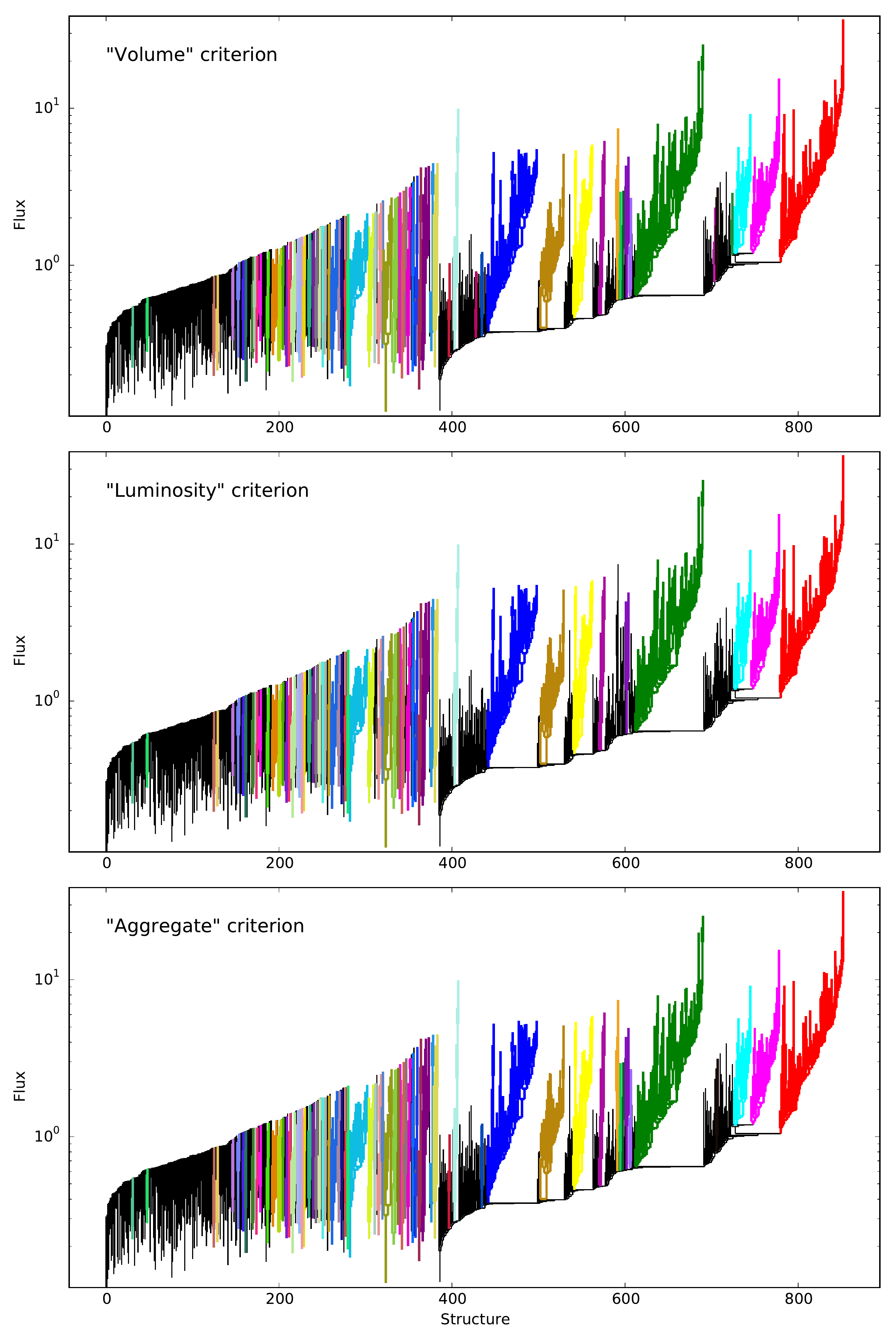}}
\caption{Dendrogram of the Orion-Monoceros complex obtained using the same parameters as in figure~\ref{F:orion_affmats} through (from the top to the bottom) the ``volume'', ``luminosity'' and ``aggregate'' criteria, respectively. Every color region outlines structures belonging to a certain cloud as segmented by SCIMES.}
\label{F:orion_dendro}
\end{figure*}

Although the above choices of similarity might be one of the most reliable criteria to constrain both $\sigma_{S}$ and $k$, it is not free from ambiguities.  Moreover, running the spectral clustering for all possible clustering configurations and for a quite large number of scaling parameter is computationally expensive. Nevertheless, there is a natural way to efficiently select $\sigma_{S}$ that might be of great interest in the segmentation of GMCs.  From a similarity matrix we can build a \emph{similarity histogram}. If the data form clusters, the histogram for their similarities will be multi-modal (figure~\ref{F:orion_amshisto}). In this view, the first mode corresponds to the average intra-cluster similarity, while the others to similarities between-clusters (\citealt{fisher_poland04}). By choosing the scaling parameter between the first two modes, the similarity values for the leaves forming clusters or single clouds, are expected to be enhanced compared to the others. We note that this choice produces regular block diagonal matrices. Therefore, a good choice of $\sigma_{S}$ that does not underestimate or overestimate the size of the clusters is between the first and the second mode. For the algorithm, we use their median value.  Physically the scaling parameter picked in this way might indicate the characteristic maximal values of volume or luminosity that the clouds tend to exhibit.

The criteria chosen to cluster the dendrogram are monotonic and produce very regular block diagonal matrices. Having rescaled the affinity matrices with an educated guess of $\sigma_{S}$, the blocks that might be related to the final clusters stand out (see figure~\ref{F:orion_affmats}, lower row) and the other affinity fall below 0.2. \texttt{SCIMES} automatically counts the equal-size squares along the main diagonal after flagging all affinity $<0.2$ and producing the starting value of the cluster number $k_{g}$. This operation is conceptually similar to the counting of the connected components of the Laplacian matrix through the Fielder vector (\citealt{fiedler73}). The Fielder vector is the first eigenvector of the Laplacian matrix and, given the properties of the Laplacian (see Section~\ref{SSS:spclust_simmat}), the zeroth elements of it represents the graph connected components if the graph is not fully connected. This method is more general than the one adopted by \texttt{SCIMES}. However, we observed that, for the chosen criteria, the simple counting of blocks gives $k_{g}$ closer to the best cluster number defined by the silhouette.

\begin{figure}
\begin{center}
\includegraphics[width=0.4\textwidth]{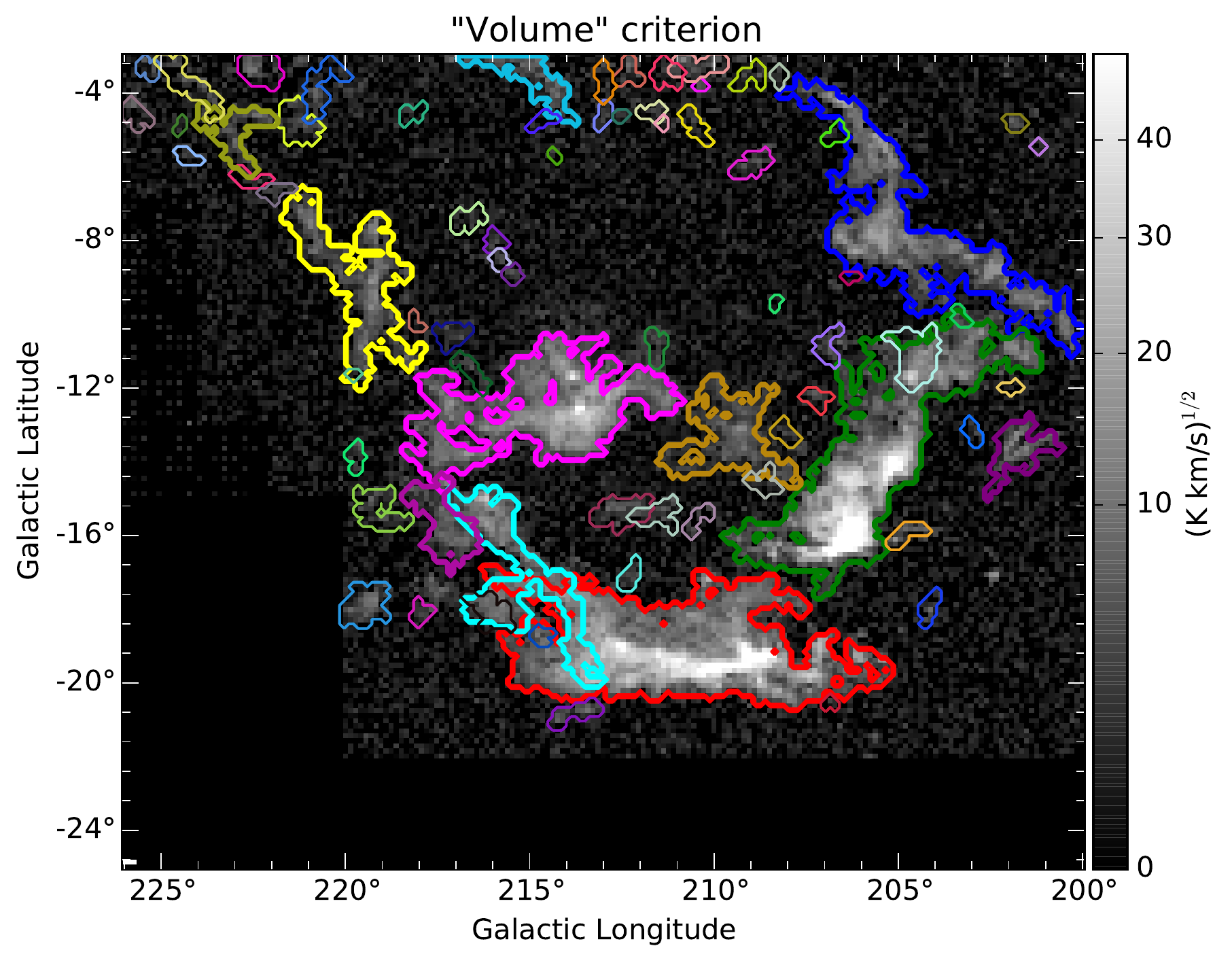}
\includegraphics[width=0.4\textwidth]{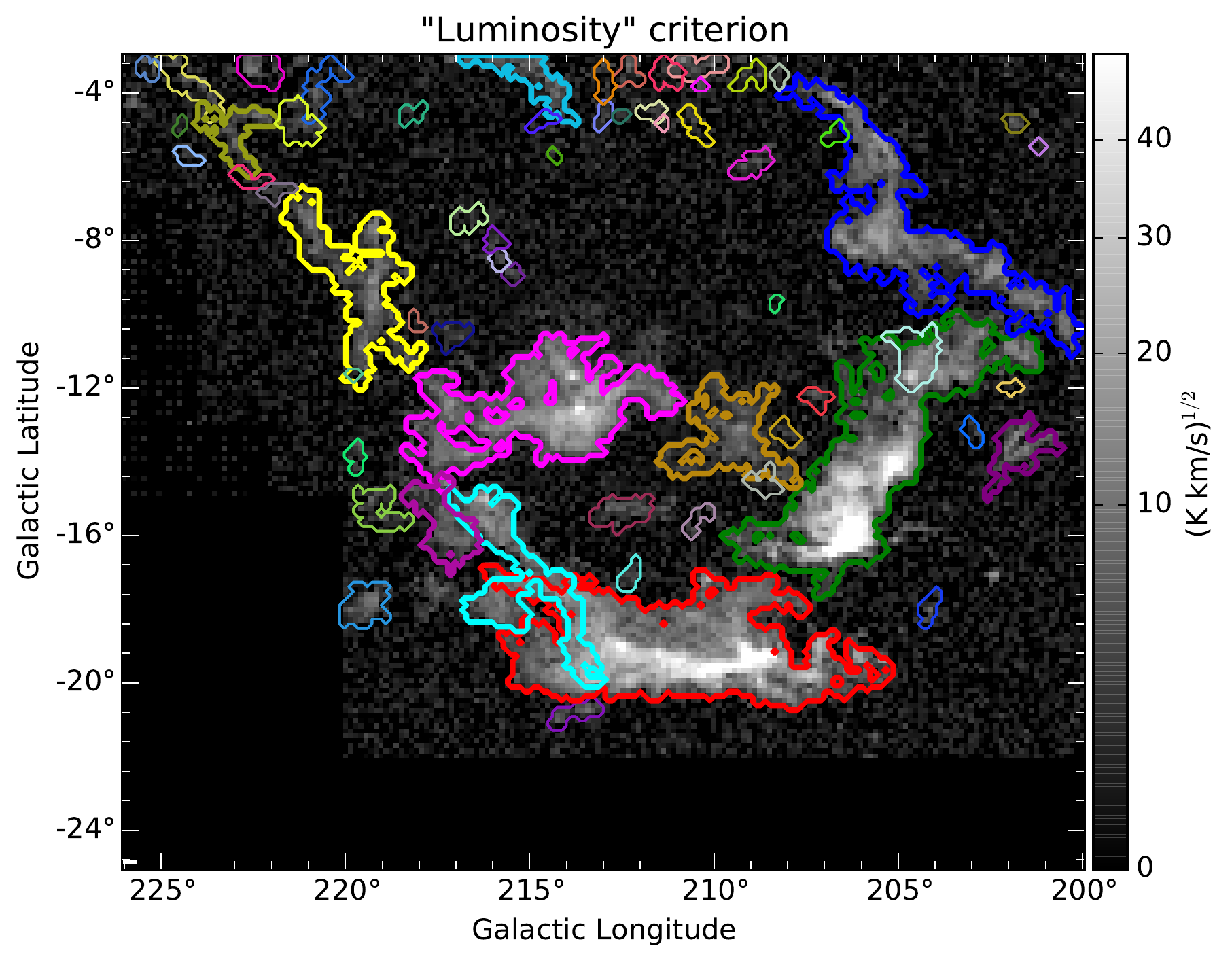}
\includegraphics[width=0.4\textwidth]{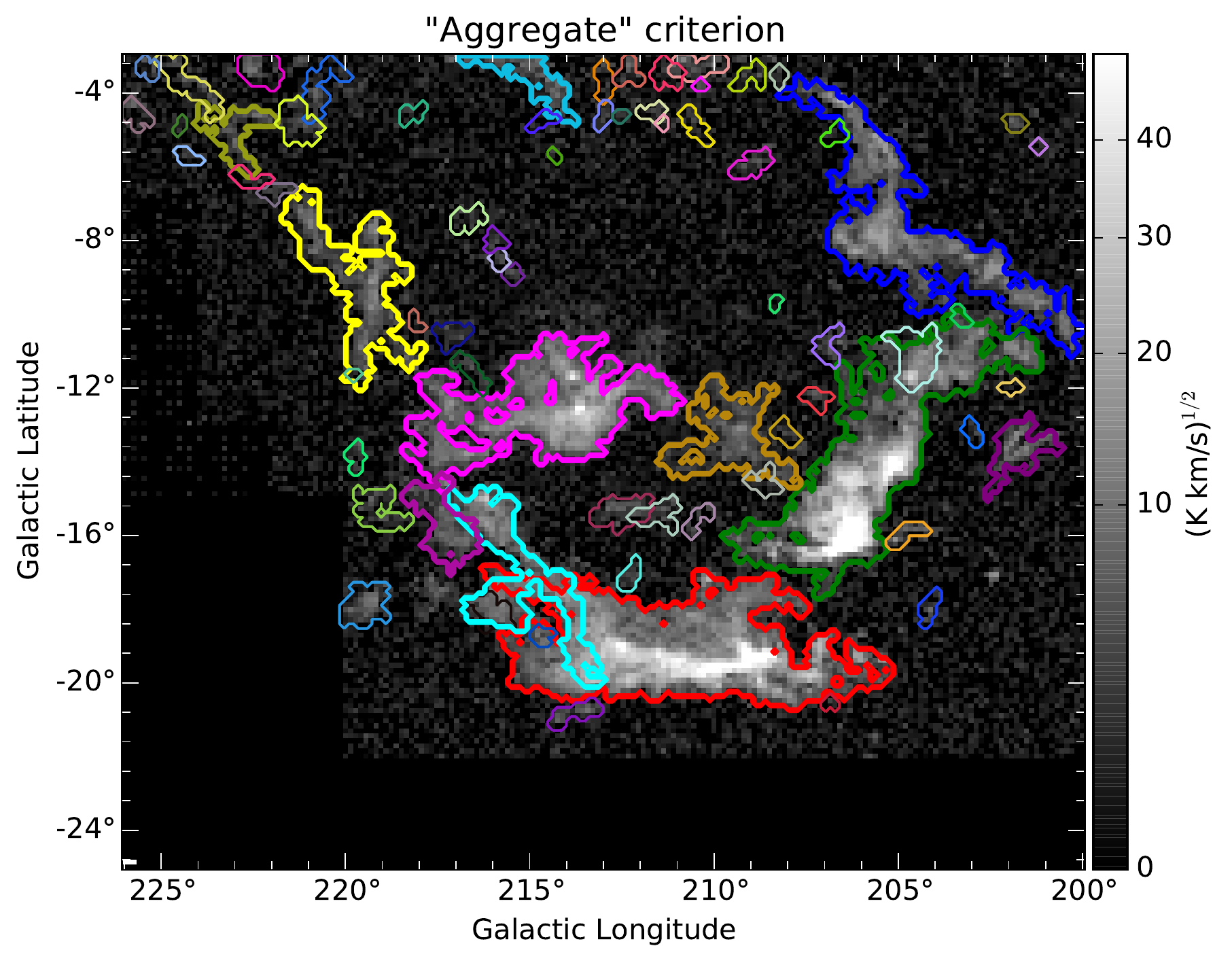}
\end{center}
\caption{The Orion-Monoceros complex square root of the integrated intensity maps. Every different contour color indicates a single cloud of the complex identified by \texttt{SCIMES}, through (from top to bottom) the ``volume'', ``luminosity'', and the ``aggregated'' criteria, respectively (see Section~\ref{S:scimes} for details). The contours use the same color scheme as figure~\ref{F:orion_dendro} and encompass isosurfaces corresponding to the clustered dendrogram branches of that figure.}
\label{F:orion_cont}
\end{figure}

%---------------------------------------------------------------------
\subsection{Cluster removal and final cloud identification}
\label{SS:scimes_rem} 
%---------------------------------------------------------------------
\noindent The most appropriate number of clusters, $k_{g}$, is guessed using the similarity matrix histogram (section \ref{SS:scimes_sigma}, see also figure~\ref{F:orion_amshisto}). Then, the code runs for several values around $k_{g}$ and the $k$ value corresponding to the highest average silhouette is selected as the best number of clusters $k_b$ (Section \ref{SSS:spclust_clnum}). Spectral clustering, by definition, groups all objects into different clusters. Leaves that do not form isolate clusters, are grouped all together in sparse clusters without any neighbors in PPV space between constituent objects. These leaves are located and eliminated from the clustering labels.  The final clouds are branches that contain only leaves in a single cluster. These clouds are, therefore, structures already considered by the original dendrogram algorithm and constitute the relevant objects embedded within the dendrogram. The application of the kernel, with a specific scaling parameter, to the similarity matrix enhances the similarity of pair of leaves above critical values for $L$ and $V$ and drastically reduces the others. Accordingly the selected clouds would tend to present similar properties in terms of luminosity and volume.  This implies that the clouds are found at different hierarchical level but with similar characteristic properties.  

%=====================================================================
\section{Testing the method}
\label{S:test} 
%=====================================================================

In this section, we apply \texttt{SCIMES} to the Orion-Monoceros complex. This system is one of the most studied star formation regions in the Galaxy with a well established set of clouds with a molecular mass $\geq 10^{4}$\,M$_{\odot}$ (\citealt{wilson05}, table 2). It represents, therefore, an ideal testbed for the capabilities of the algorithm.

%---------------------------------------------------------------------
\subsection{The Orion-Monoceros CO(1-0) dataset segmentation}
\label{SS:test_oriondata} 
%---------------------------------------------------------------------

The Orion-Monoceros complex dataset we use in our tests has been presented by \citep[][figure~\ref{F:orion_cont}]{wilson05}. The $^{12}$CO(1-0) data were obtained with the 1.2m millimeter wave telescope at the Harvard-Smithsonian Center for Astrophysics and present a spatial resolution of $\theta_{\mathrm{FWHM}}=8.4$ arcminutes corresponding to $\sim1$\;pc at the average distance to the complex ($\sim450$\;pc). The field-of-view spans a region of $\sim26^{\circ}\times19^{\circ}$ or $\sim200$\;pc$\times160$\;pc. The data cube has a velocity resolution of 0.65\;km\;s$^{-1}$, over a $v_{LSR}$ range between $-3$ to 19.5\;km\;s$^{-1}$.  However, most of the complex molecular emission is concentrated between 2 and 15\;km\;s$^{-1}$. The data have a roughly uniform sensitivity of $\sigma_{\mathrm{rms}}=0.26$\;K.\\

\begin{figure*}
{\includegraphics[width=0.85\textwidth]{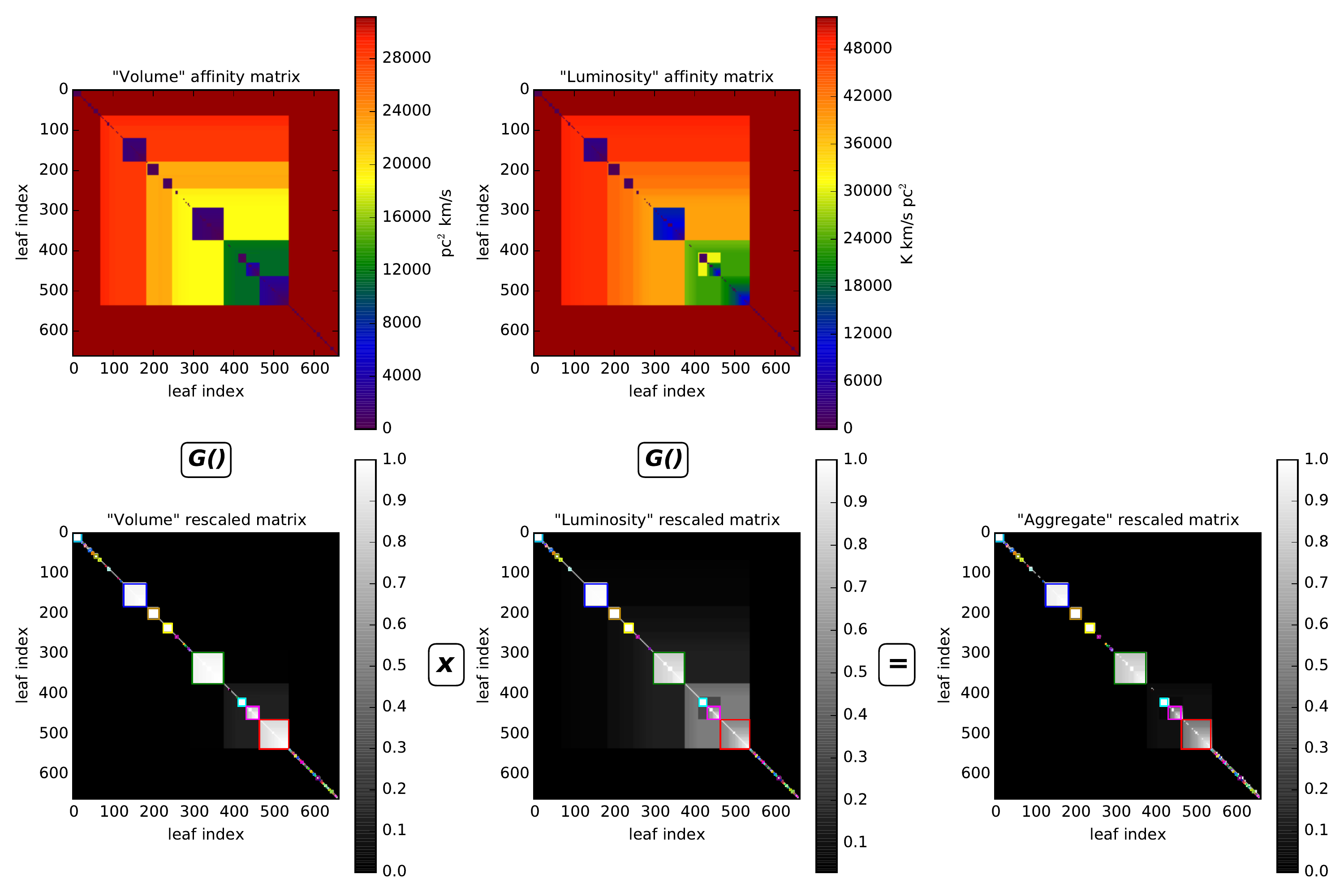}}
\caption{Similarity matrices obtained from the Orion-Monoceros complex dendrogram. For this example, the \texttt{astrodendro} parameters have been set as in Section~\ref{SS:test_oriondata}. The first row indicates the original similarity matrices associated with the ``volume'' and ``luminosity'' criteria, $S_{V}$ and $S_{L}$, respectively; the second row shows the kernel rescaled matrices, $G(S_{V}) = \exp(-S_{V}^{2}/\sigma_{S,V}^{2})$ and $G(S_{L}) = \exp(-S_{L}^{2}/\sigma_{S,L}^{2})$; in the last (lower right) panel, the ``aggregate'' similarity matrix, given by the element-by-element multiplication of $G(S_{V})$ and $G(S_{L})$ is shown. The matrices rows and columns are labeled with the dendrogram leaf indexes. Within the different panels the operations between the different matrices are indicated, where $G$ is the rescaling with the kernel. Matrix cells corresponding to leaves of a certain cluster are contoured using the same color as in figure ~\ref{F:orion_dendro}.}
\label{F:orion_affmats}
\end{figure*}

Figure~\ref{F:orion_dendro} shows the dendrogram of the dataset obtained with typical parameter values (\verb"min_delta"$=2\sigma_{\mathrm{rms}}$, \verb"min_npix"$=3\theta_{\mathrm{FWHM}}\sim3.6$ pixels). The \verb"min_value" has been set to zero, since the datacube has been previously masked using the \emph{dilate mask} approach (\citealt{rl06}). The technique works by masking pixels in two consecutive velocity channels in which the signal is above $4\sigma_{\mathrm{rms}}$. These regions are then extended to include all adjacent pixels in which the signal is above $1.5\sigma_{\mathrm{rms}}$. The rms noise $\sigma_{\mathrm{rms}}$ of the Gaussian distribution is estimated from the outlier-robust median absolute deviation (MAD) of each spectrum. In this way, we retain most of the significant emission within the data cube, even when the noise is not spatially homogeneous.  A catalog of each dendrogram structure has been generated using the \verb"astrodendro" package \verb"ppv_catalog" method, which measures moment-based properties at a set of levels in the dendrogram.  Further, the similarity matrices of figure~\ref{F:orion_affmats} (see Section~\ref{SSS:spclust_simmat}) have been obtained using the criteria in Section~\ref{SSS:scimes_fluxvol}. To convert in physical units we use the distances of \citet[][their Table 2]{wilson05}. \texttt{SCIMES} identified the scaling parameter $\sigma_{S,V}\sim7940$\,pc$^2$\,km\,s$^{-1}$ and $\sigma_{S,Lum}\sim28128$\,K\,km\,s$^{-1}$\,pc$^2$ for volume and luminosity criteria, respectively that have been used to rescale the matrices through the kernel.  In order to make a comparison of different measurements, we ran \texttt{SCIMES} on the volume, luminosity, and aggregate criteria separately. The similarity matrices relative to these criteria are shown in the last row of figure~\ref{F:orion_affmats}. Direct analysis of the similarity matrices predicted cluster numbers $k_{g}=\{74, 69, 74\}$ for the volume, luminosity and aggregate criteria respectively. Silhouette values equal to $\{ 0.97,0.86,0.94\}$, however, identifies more appropriate cluster numbers $k_{b}=\{76,61,70\}$. According to the criteria in Section~\ref{SS:scimes_rem}, three clusters have been removed from the volume-based segmentation, however no clusters have been removed from the luminosity and aggregate criteria-based segmentations. The final dendrogram decompositions are shown in figure~\ref{F:orion_dendro}, while the corresponding maps of the objects identified by \texttt{SCIMES}, using the same color scheme of the dendrograms, in figure~\ref{F:orion_cont}.\\

We immediately note that the canonical clouds of the Orion-Monoceros complex (i.e. Orion A - red contour, Orion B - green contour, the Northen Filament - blue contour, Monoceros - magenta contour, the Crossbones - yellow contour, NGC2149 - cyan contour, the Scissor - purple contour) are always faithfully recognized by the algorithm as single entities providing segmentation close to a ``by-eye'' work. Other smaller objects are included only if they encompass at least two leaves.

%---------------------------------------------------------------------
\subsection{Difference between volume, luminosity and aggregate criteria segmentations}
\label{SS:test_vollum} 
%---------------------------------------------------------------------

\noindent Figures~\ref{F:orion_cont} show the different Orion-Monoceros complex segmentations provided by volume, luminosity and aggregate criteria applied separately. In general, all notable clouds in the complex are retained regardless of which similarity criterion we use.  A few additional, smaller objects are missed by the luminosity- and aggregate-based decompositions. By looking at the rescaled matrices in figure~\ref{F:orion_affmats}, the clouds of the complex appear as well-defined square sub-matrices within the main for the volume criterion. This matrix is therefore the dominant one once we aggregate the two basic criteria. Indeed the silhouette value provided by the volume clustering configuration is very high (0.97) and higher than the values obtained using the other criteria (0.86 and 0.94 for luminosity and aggregate criteria respectively). Therefore, for the following tests and analysis we will only consider the segmentation provided by the volume criterion.

%---------------------------------------------------------------------
\subsection{Algorithm stability with the noise}
\label{SS:test_noise} 
%---------------------------------------------------------------------

\begin{figure*}
\centering
{\includegraphics[width=1\textwidth]{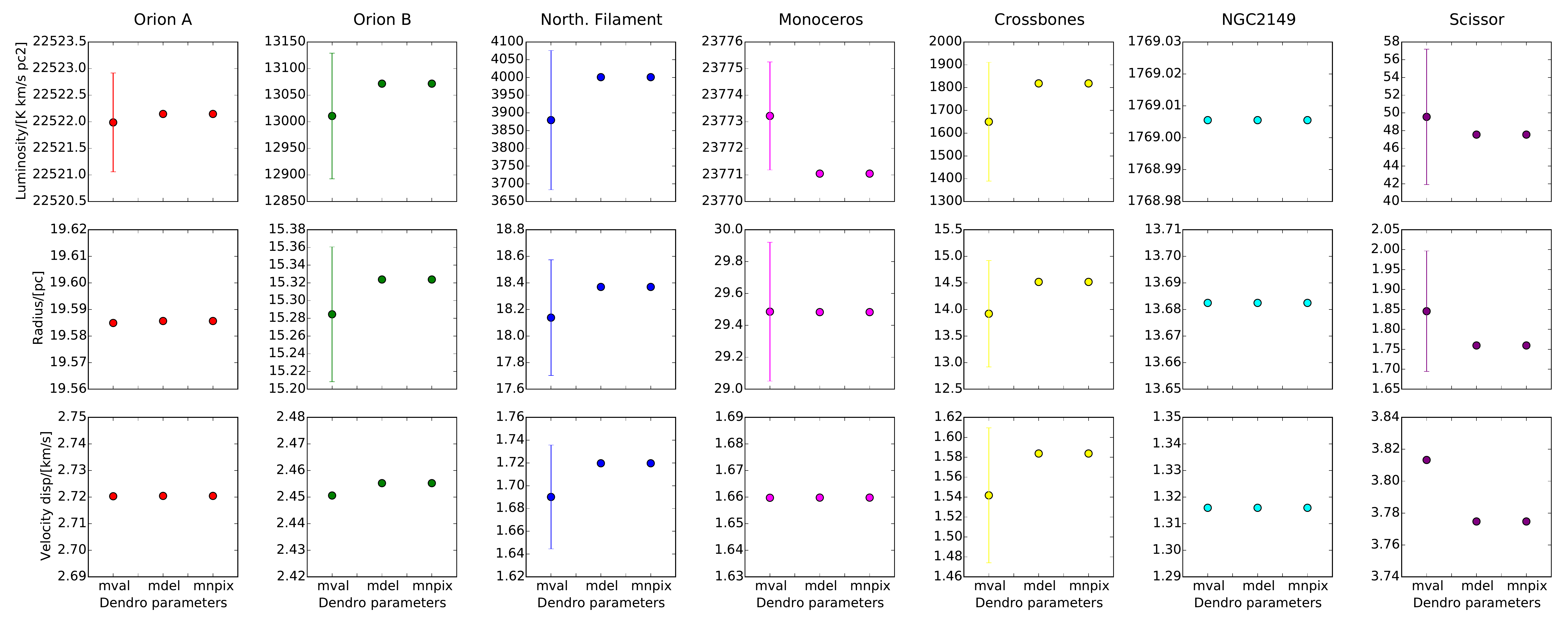}}
\caption{Results on the SCIMES robustness with variations of dendrogram parameters on the Orion-Monoceros complex canonical clouds.  The cloud properties are shown on the $y$-axis from top to bottom: CO luminosity, radius, velocity dispersion. The $x$-axis indicates which parameter varies for a given test as exposed in the text, where mval = min\,value, mdel = min\,delta, mnpix = min\,npix. Error bars indicate the amount of variation as standard deviation of a certain cloud property with the dendrogram parameter. The absence of an error bar indicates a variation equal to zero (no variation of the cloud property with changes of the given dendrogram parameter) or that the variation is smaller than the size of the symbol.}
\label{F:orion_robust}
\end{figure*}

%sigmas = [8316.62389631,8872.84887998,7559.70009544,5532.48567496,9019.80216308,8027.67076641,7869.17070411,4903.38805498,8449.36360461,7509.92296167]

\noindent To test the algorithm stability in the presence of noise and its ability to provide reliable results down to the noise level, we add  beam-correlated noise to the Orion-Monoceros datacube.  We analyze the properties of the notable clouds imposing the typical parameters (see Section~\ref{SS:test_oriondata}) and \verb"min_value" $=1\sigma_{\mathrm{rms}}\sim0.36$\,K for the generation of the dendrogram. We generate 10 datacubes with random noise realizations. We use \texttt{SCIMES} with the default settings and we perform the decomposition based on the volume matrix only. For all noise-added cubes, the volume matrix shows scaling parameters that are quite stable: $\sigma_{S} = 7600\pm1300$\,pc$^{2}$\,km\,s$^{-1}$. Table~\ref{T:noisereal} reports the results of the test as mean value and standard deviation of the cloud properties measured for the different cubes. Generally, \texttt{SCIMES} behaves well, yielding cloud properties that do not vary significantly with the different noise realizations: properties differ by only few percent up to a maximum of 15\% between the cubes.  Nevertheless, adding random noise alters the significance of the local maxima of the original datacube, and consequently the connectivity of the various objects resulting in some changes in the cloud identification.   

\begin{table}
\begin{tabular}{cccc}
\hline
Object & CO luminosity & Radius & Velocity dispersion \\
& K\,km\,s$^{-1}$\,pc$^{2}$ & pc & km\,s$^{-1}$ \\
\hline
Orion A & $22871\pm451$ & $19.7\pm0.3$ & $2.7\pm0.0$ \\
Orion B & $13144\pm295$ & $15.5\pm0.3$ & $2.5\pm0.0$ \\
North. Fil. & $4048\pm182$ & $18.3\pm0.3$ & $1.8\pm0.1$ \\
Monoceros & $21912\pm3708$ & $27.3\pm3.3$ & $1.6\pm0.1$ \\
Crossbones & $1759\pm228$ & $14.1\pm0.7$ & $1.6\pm0.1$ \\
NGC2149 & $1845\pm271$ & $13.5\pm1.4$ & $1.4\pm0.1$ \\
Scissors & $66\pm2$ & $2.3\pm0.2$ & $3.9\pm0.1$ \\
\hline
\end{tabular}
\caption{Properties of the most notable Orion-Monoceros complex clouds. CO luminosity, radius, and velocity dispersion are indicated as mean and standard deviation of the properties measured of the ten datacubes with different noise realizations.}
\label{T:noisereal}
\end{table}

%---------------------------------------------------------------------
\subsection{Algorithm robustness with the dendrogram parameters}
\label{SS:test_robustpar} 
%---------------------------------------------------------------------

In this section, we study the performance of the algorithm to provide robust results with variations of the dendrogram parameters \verb"min_value", \verb"min_delta", \verb"min_npix".
We perform three tests varying a single parameter and fixing the others to typical values:

\begin{enumerate}

\item \verb"min_value" $=\{1,1.5,2,2.5,3\}\sigma_{\mathrm{rms}}$, \verb"min_delta" $=2\sigma_{\mathrm{rms}}$, \verb"min_npix" $=3\theta_{FWHM}$;

\item \verb"min_delta" $=\{1,2,3,4,5\}\sigma_{\mathrm{rms}}$, \verb"min_value" $=2\sigma_{\mathrm{rms}}$, \verb"min_npix" $=3\theta_{FWHM}$;

\item \verb"min_npix" $=\{1,2,3,4,5\}\theta_{FWHM}$, \verb"min_value" $=2\sigma_{\mathrm{rms}}$, \verb"min_delta" $=2\sigma_{\mathrm{rms}}$.

\end{enumerate}

\noindent Figure~\ref{F:orion_robust} shows the result of the tests as variations of CO luminosity, radius, velocity dispersion of Orion-Monoceros complex most notable objects (Orion A, Orion B, the Northen Filament, Monoceros, the Crossbones, NGC2149, the Scissors) with changes on the dendrogram parameters. Overall, it appears that \emph{SCIMES is robust against the parameters used to define the starting dendrogram}. Cloud properties are more sensitive by changes of \verb"min_value", i.e. the noise level at which the dendrogram has been generated. Nevertheless, those differences do not concern all objects and all properties. The clouds more affected by variations in \verb"min_value" are the smaller ones (e.g. the Crossbones and the Scissors), the properties of which vary by 10-20\% at most. For the larger GMCs as Orion A, the Northern Filament and Monoceros, differences in the properties are irrelevant. All objects appear insensitive to variations  in the other two dendrogram parameters \verb"min_delta" and \verb"min_npix". In particular, NGC2149 is not affected by any parameter variation.   Overall, the dendrogram parameters have a minimal effect on segmentation outcomes.

\begin{figure*}
\centering
\begin{tabular}{cc}
\includegraphics[width=0.4\textwidth]{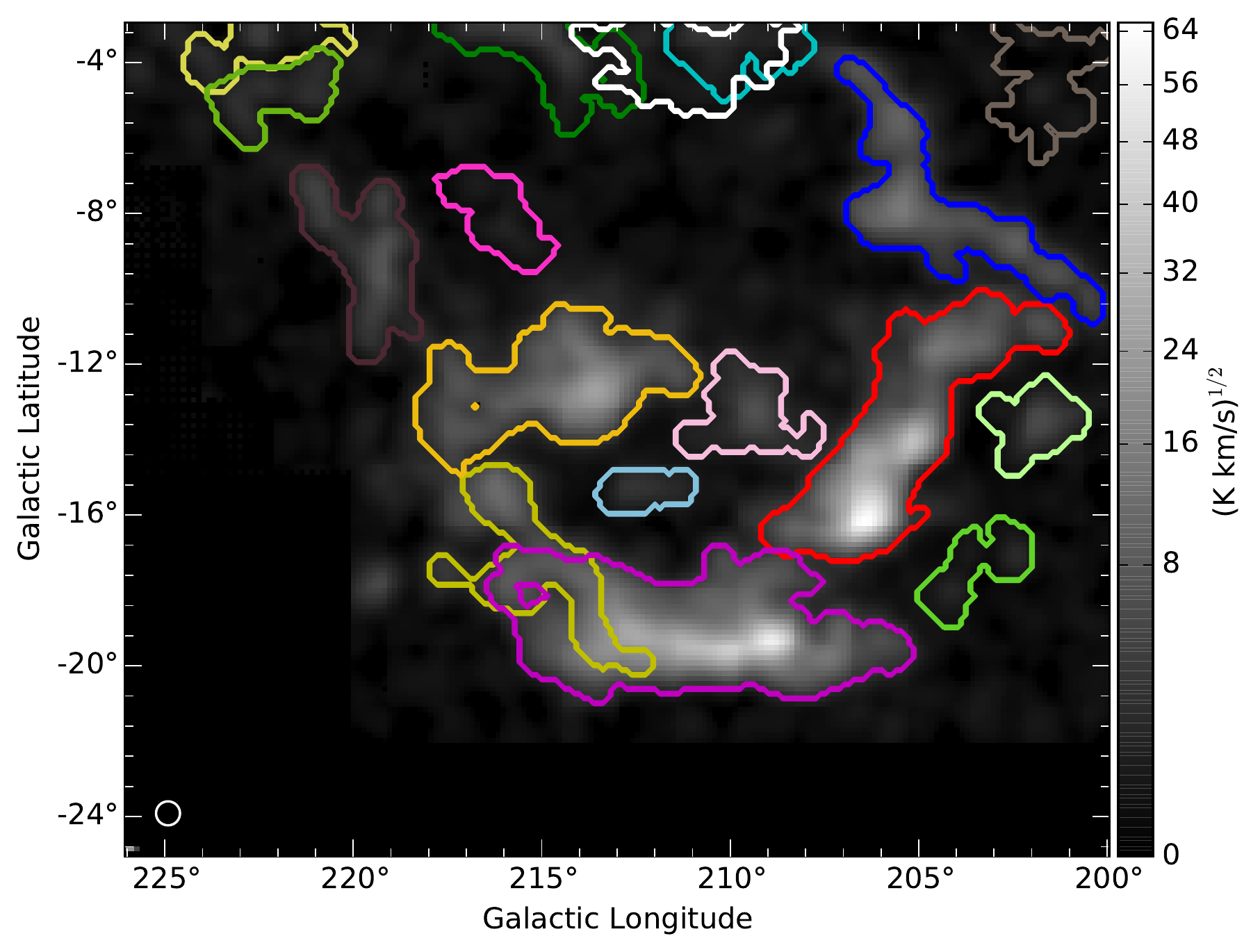} &
\includegraphics[width=0.4\textwidth]{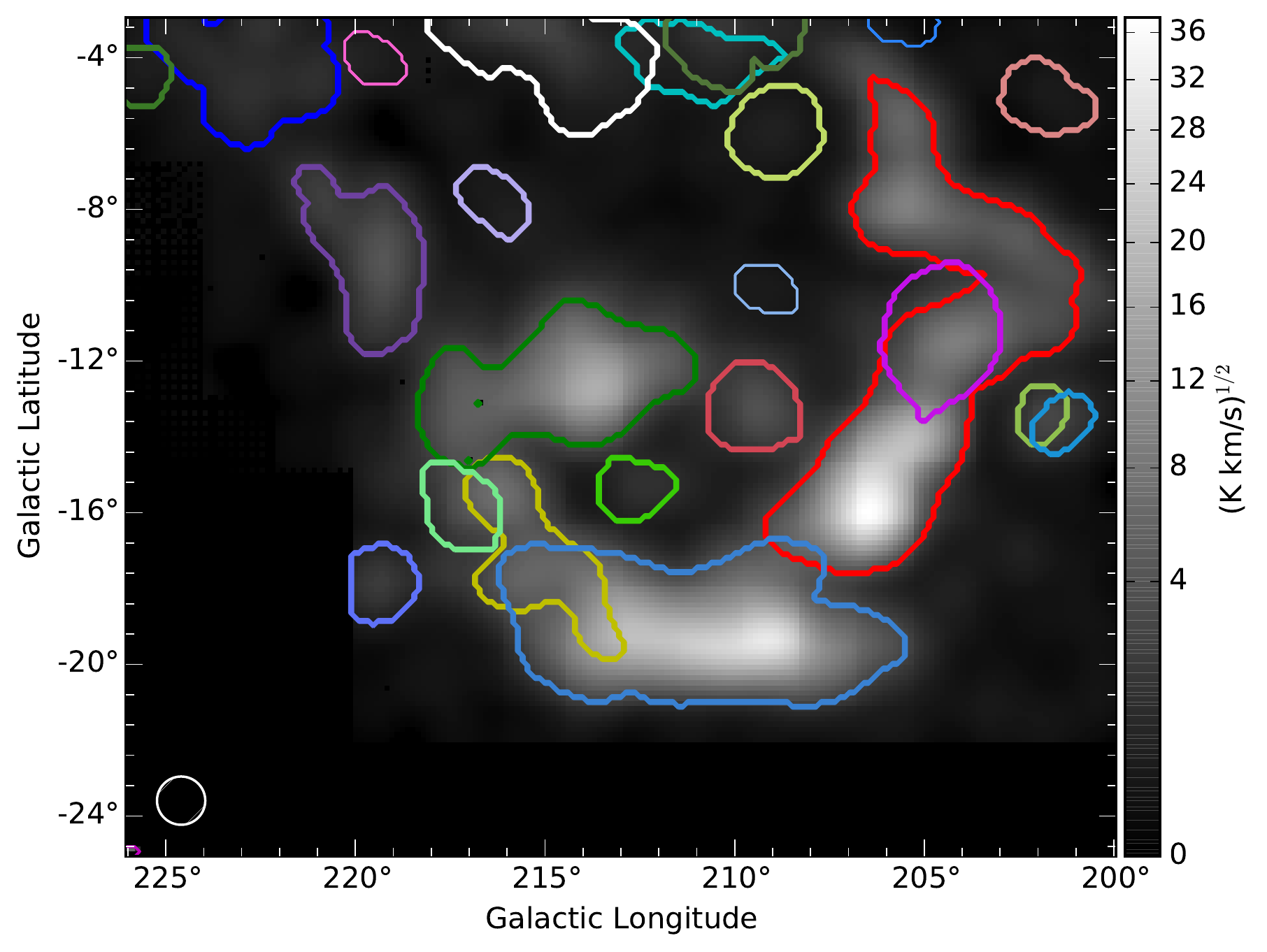} \\
\includegraphics[width=0.4\textwidth]{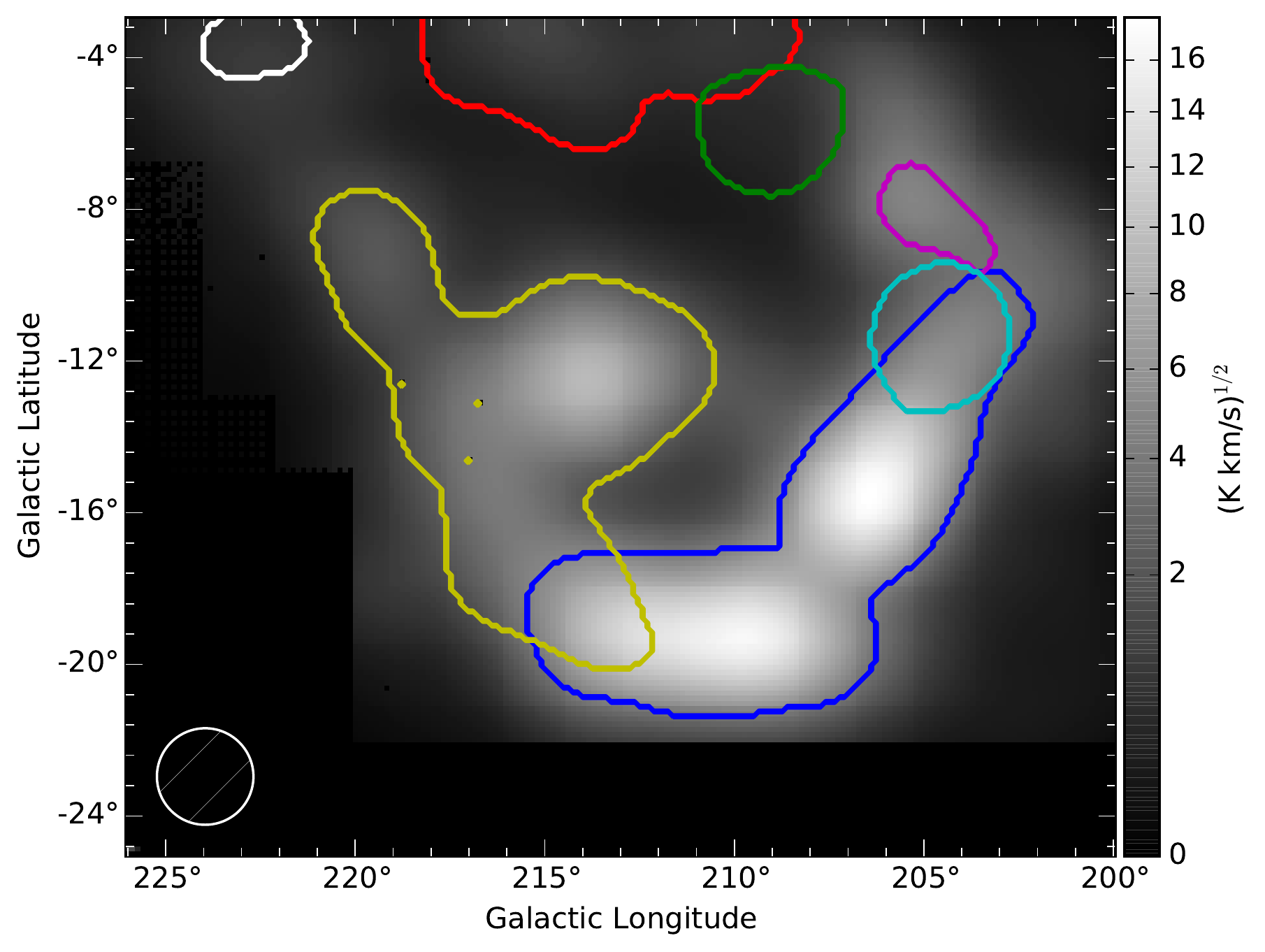} &
\includegraphics[width=0.4\textwidth]{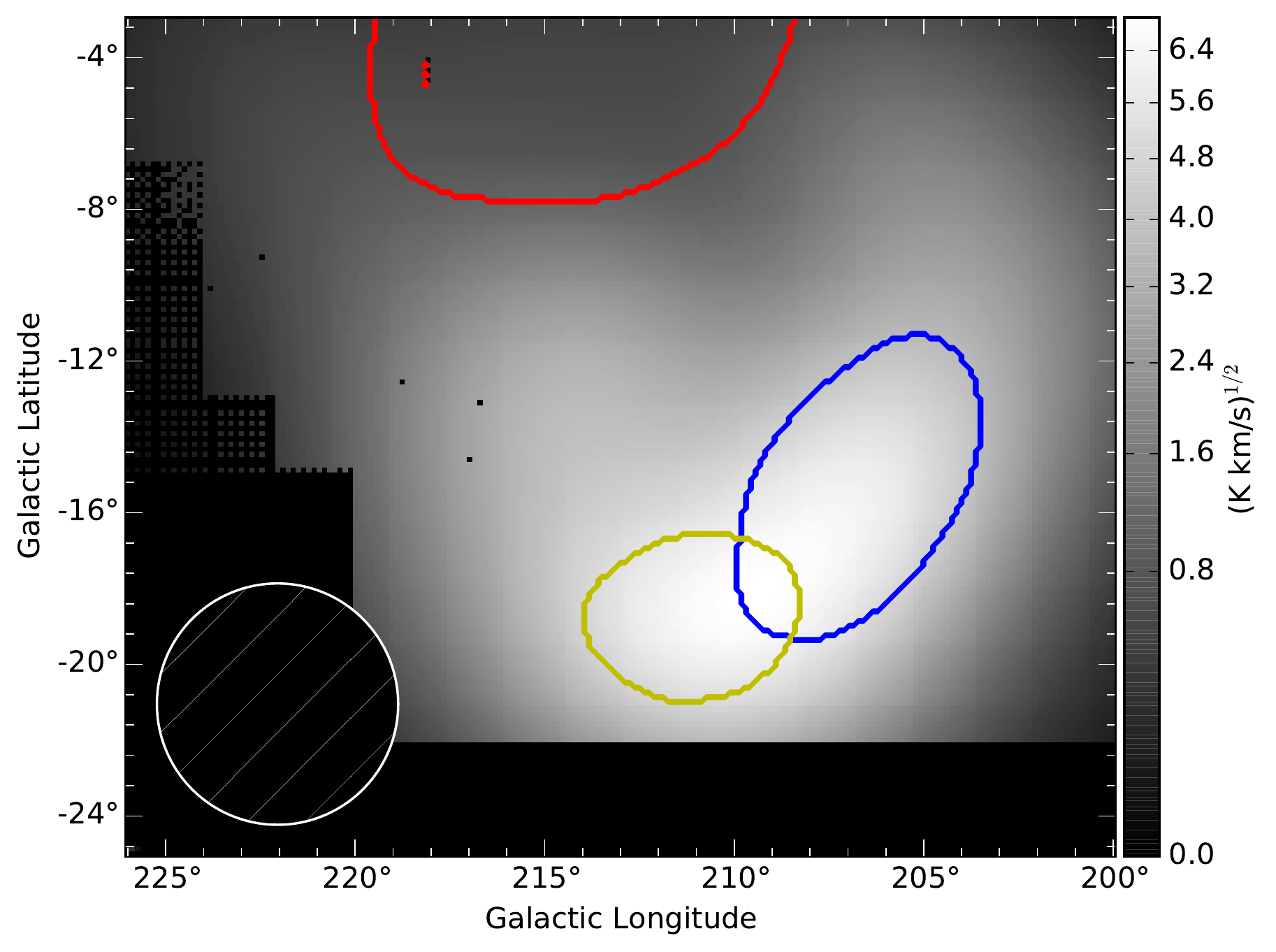} \\
\end{tabular}
\caption{Orion-Monoceros complex cloud segmentation at different spatial resolution; from left to the right, top to bottom: 5\,pc, 10\,pc, 20\,pc, 50\,pc. The beam is indicated at the lower left corner of each panel.}
\label{F:orion_res}
\end{figure*}

%---------------------------------------------------------------------
\subsection{Algorithm performance at lower resolution}
\label{SS:test_res} 
%---------------------------------------------------------------------

\noindent To verify the behavior of \texttt{SCIMES} at lower resolutions we smooth the Orion-Monoceros dataset to 5\,pc, 10\,pc, 20\,pc, 50\,pc spatial resolution, considering an average distance to the complex of 450\,pc and using a round Gaussian kernel. Dendrograms of each smoothed datacube are generated using \verb"min_value" $=2\sigma_{\mathrm{rms}}$, \verb"min_delta" $=2\sigma_{\mathrm{rms}}$, \verb"min_npix" $=1\theta_{FWHM}$. Fig~\ref{F:orion_res} shows the result of the segmentations performed through the ``volume'' matrix only. In terms of identification, the results appear stable at 5\,pc resolution (i.e. almost 5 times lower resolution than the original dataset): the clouds are recognized as the original dataset (see Figure~\ref{F:orion_dendro}), with 30\% maximum difference between their properties. At this resolution, the scaling parameter estimated by the code ($\sigma_{S; \mathrm{5\,pc}} = 8230$\,pc$^2$\,km\,s$^{-1}$) is very similar to the one obtained from the original dataset ($\sigma_{S; \mathrm{1\,pc}} = 7940$\,pc$^2$\,km\,s$^{-1}$). At 10\,pc, Orion B and the Northern Filament are merged into the same object. This is because the Northern Filament is a single leaf at this resolution. ``Clustering'' means to group objects together that have similar properties. A single object (in our case, a ``stray'' leaf) is not, by definition, a cluster. For the same reason, using the default settings of the algorithm, several important objects (like Orion A) are missed by the decomposition. To retain those clouds, we configure the algorithm to retain single leaves within the list of clusters. This method is therefore important when the beam size becomes closer to the physical size of the GMCs as in most extragalactic observations. The scaling parameter estimated by the algorithm ($\sigma_{S; 10\,pc} = 7990$\,pc$^2$\,km\,s$^{-1}$) to rescale the volume affinity matrix is again similar to the one obtained at native resolution. However, applying a lower scaling parameter ($\sigma_{S}\sim 4000$\,pc$^2$\,km\,s$^{-1}$) allows for the segmentation of the Northern Filament separated from Orion A. At 20\,pc all notable clouds are single leaves. The clustering merges NGC\,2149, Monoceros, the Crossbones and Orion B, Orion A into two separated clusters. The estimated scaling parameter is $\sigma_{S; 20\,pc} = 4330$\,pc$^2$\,km\,s$^{-1}$. At this resolution, therefore, \texttt{SCIMES} becomes less efficient and the stability of the results compared with the dataset at the original resolution is strongly reduced. At 50\,pc resolution, the dendrogram is a single branch of two leaves and another stray leaf. In this case, \texttt{SCIMES} does not make any attempt to cluster the dendrogram, considering all leaves as separated objects. Here only Orion A and Orion B are recognized.  We conclude that a physical resolution of 10 pc or better is required to use \texttt{SCIMES} to find GMCs.  However, this regime is where other algorithms struggle to find clouds (section \ref{S:algo_comp}). At poorer resolutions, clouds are effectively point sources and are better identified by other finding routines (as CPROPS or CLUMPFIND).

%---------------------------------------------------------------------
\subsection{Nature of the unclustered emission}
\label{SS:unclust} 
%---------------------------------------------------------------------

\begin{figure}
\centering
\includegraphics[width=0.4\textwidth]{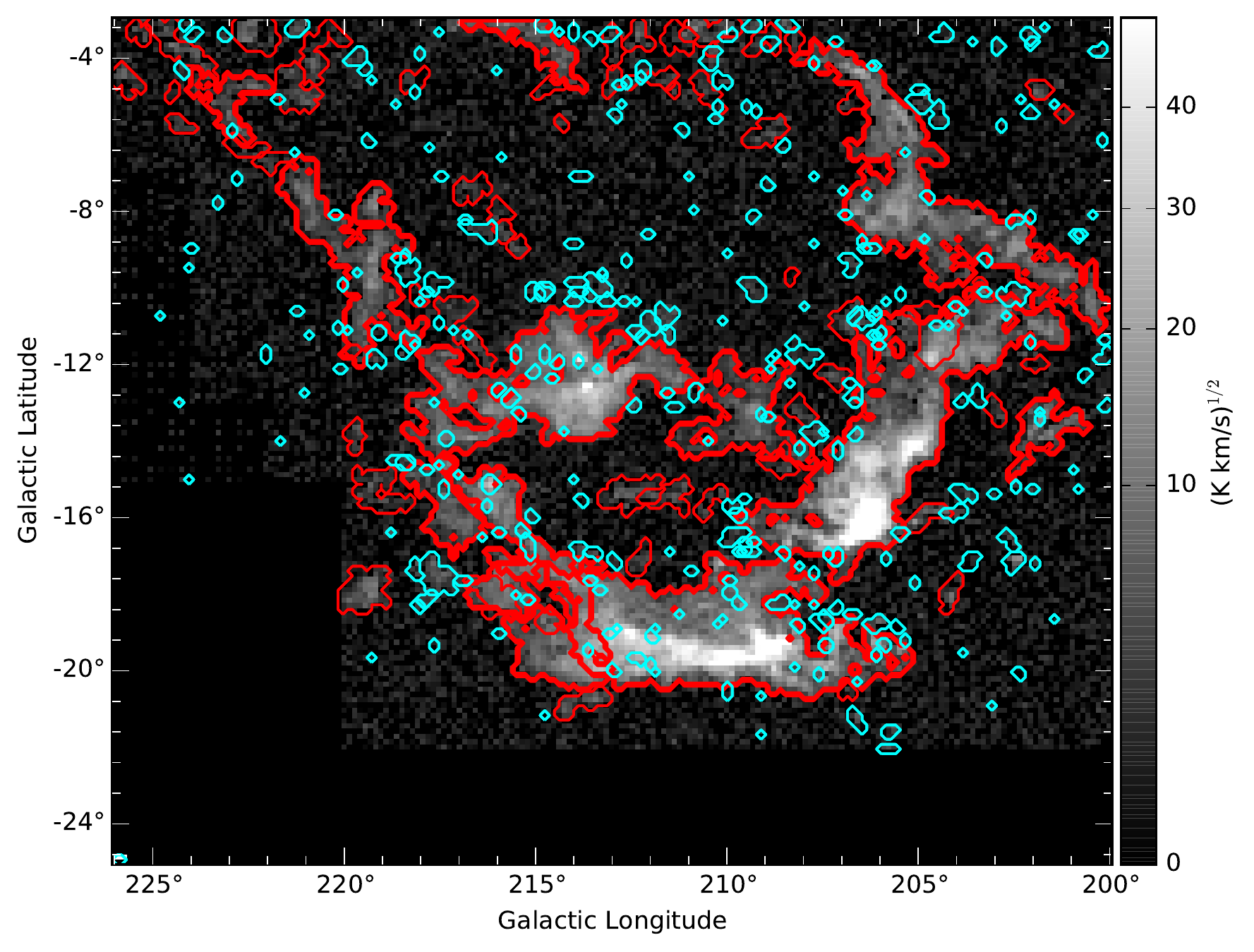}
\caption{Distribution of the unclustered structures (leaves of the dendrogram, cyan contours) and cataloged clusters (red contours).}
\label{F:orion_unclust}
\end{figure}

As described in Section~\ref{SS:scimes_rem}, \texttt{SCIMES} does not consider dendrogram leaves that cannot be uniquely attributed to separate branches. Those branches will eventually constitute single clusters of the star forming complex under consideration. Moreover, isolated leaves connected with the other structures of the dendrogram only through the artificial ``super-structure'' called trunk (see section~\ref{SS:dendro_graph}) are also eliminated from the catalog, since they are not clusters by definition\footnote{A cluster is a group of objects, in our abstraction, a group of leaves connected because of some underlying criteria. Therefore a single, isolated leaf can not be considered as a cluster.}. Those unclustered structures are colored in black in Figure~\ref{F:orion_dendro}. Nevertheless those emission structures might be significant (since they should be at least 2$\sigma_{\mathrm{rms}}$ from the merge level with another structure, however this criterion is not valid for truly isolated leaves) and well resolved (spanning at least 3 beam sizes and 2 channel width).  Thus, we examine their properties here.  Figure~\ref{F:orion_unclust} shows the distribution of these structures with respect to the cataloged clusters in the Orion-Monoceros complex. In general, the unclustered emission appears homogeneously distributed around the main objects, both in spatial and spectral sense. Those structures have an average radius of 1\,pc, and a typical velocity dispersion $\sim1$\,km\,s$^{-1}$. Moreover they encompass only 3\% of the total flux of the dataset, while the cataloged clusters contain $\sim80\%$ of the total CO emission, independent of the criterion used for the clustering. Their significance is typically a factor 2 lower than the properties of the cataloged structures and only 5\% of them has a peak signal-to-noise ratio above 2. The latter might, therefore, be real entities of the molecular ISM, comparable to the smaller objects cataloged by \texttt{SCIMES}. In conclusion, the algorithm retain most of the significant emission of the dataset and the unclustered emission represents mostly noisy peaks rather than small structures in the molecular medium.

%=====================================================================
\section{Comparison with other cloud identification methods}
\label{S:algo_comp} 
%=====================================================================

In this Section we will compare \texttt{SCIMES} segmentation and cluster properties to those provided by other popular cloud identification algorithms. In particular we will consider the dendrogram itself, CPROPS (\citealt{rl06}) and CLUMPFIND (\citealt{williams94}). We do not include GAUSSCLUMPS in the tests, since the code fits a 3D Gaussian to the molecular emission, assuming, therefore, a defined morphology of the molecular structures. This is a rough approximation of the real shape of the molecular clouds likely suitable for extragalactic observation where the beam filling factor is generally lower than the unity. In the Galactic surveys at high resolution, instead, molecular structures show a variety of shapes that have little resemblance with Gaussians.

%---------------------------------------------------------------------
\subsection{Segmentation differences}
\label{SS:algo_comp_asgn} 
%---------------------------------------------------------------------

\begin{figure*}
\centering
\includegraphics[width=1\textwidth]{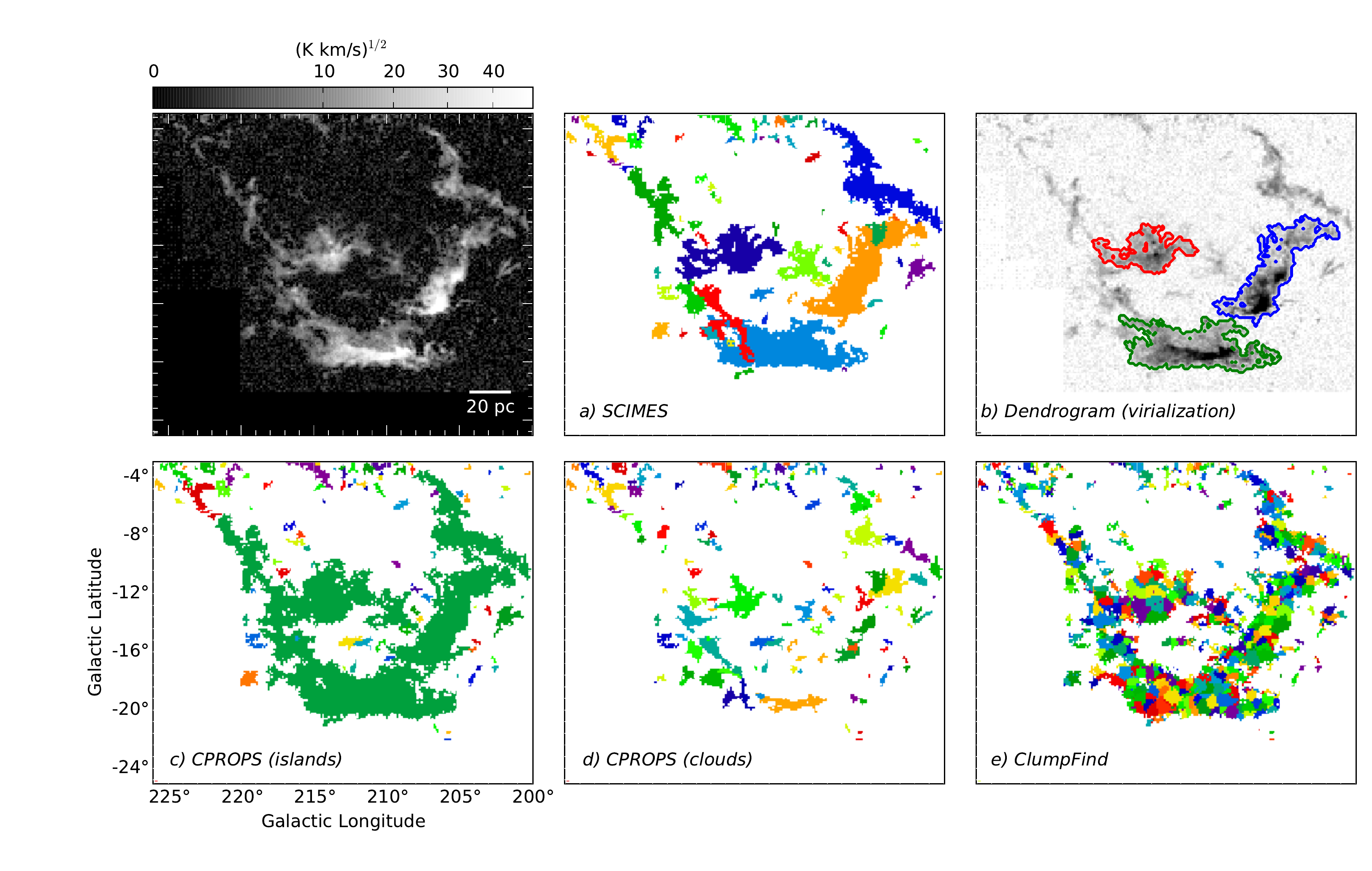}
\caption{The Orion-Monoceros complex (top left) and the emission segmentation performed by different algorithms. Decompositions are presented as collapsed assignment cubes, where each color indicates an individual objects. \emph{a)}, the cloud decomposition provided by SCIMES. \emph{b)}, the largest self-gravitating connected objects (red, blue and green contours) within the dendrogram having $1\leq\alpha<2$ as in R08. \emph{c)}, structures identified by CPROPS' island method, using the default setting of the decomposition parameters. \emph{d)}, clouds identify by CPROPS ``physical priors'' method that sets the tuning parameters to values appropriate for a GMC segmentation (see Rosolowsky \& Leroy 2006, table 2 for details). \emph{e)}, the cloud decomposition provide by CLUMPFIND, starting from the CPROPS defined islands.}
\label{F:orion_algo_comp_asgn}
\end{figure*}

\noindent Figure~\ref{F:orion_algo_comp_asgn} shows the emission segmentation of the Orion-Monoceros dataset using different algorithms.  These algorithms are tailored for different purposes and it is clear that \texttt{SCIMES} appears particularly well-suited for the cloud segmentation in data sets with high resolution. GMCs might be also identified directly from the dendrogram based on the value of the virial parameter of the emission within the isosurfaces at the various hierarchical levels. The virial parameter is a dimensionless quantity that determines the dynamical state of the clouds (\citealt{mckee_zweibel92}). It is defined as: 

\begin{equation}
\label{E:virpar}
\alpha = \frac{5\sigma_{v}^{2}R}{4.4X_{\mathrm{CO}}L_{\mathrm{CO}}G} = \frac{1.12M_{\mathrm{vir}}}{M_{\mathrm{lum}}};
\end{equation}

\noindent where $M_{\mathrm{vir}}=1040\sigma_{v}^{2}R$ and $M_{\mathrm{lum}}=4.4X_{\mathrm{CO}}L_{\mathrm{CO}}$ (\citealt{rl06}). Several studies (\citealt{solomon87}, \citealt{heyer01}) have indicated that isolated GMCs show virial parameters
$\alpha\sim1$ suggesting the self-gravitating state of those objects. In R08, the authors use the virial parameter to identify clouds in the Orion-Monoceros dataset as largest-scale self-gravitating structures within the dendrogram. In this way they obtain a good description of three clouds of the complex: Orion A, Orion B and Monoceros. Nevertheless other canonical objects are not recognized through the virial parameter approach. The virial parameter, indeed, might be not the best method to identify GMCs on different mass and size scales. The true dynamical state of the GMCs is not clear, and although on average, molecular cloud populations show virial parameters close to unity, several observations (\citealt{heyer09}; \citealt{rosolowsky07}; \citealt{bolatto08}; \citealt{colombo14a}) have indicated that a large number of clouds are unbound having $\alpha>2$. The virial parameter approach may be more useful to identify clumps within clouds that are more likely to be bound (e.g. \citealt{dib07}; \citealt{shetty10a}).

One of the most popular algorithms for the GMC decomposition is CPROPS (\citealt{rl06}). The CPROPS package includes different emission segmentation routines. The ``island'' method distinguishes as single objects connected regions of emission within the PPV space. Such approach can be sufficient to catalog discrete molecular structures in flocculent extragalactic environment, where the emission is typically sparsely distributed (e.g. the LMC; \citealt{wong11}).  However, in more complex galaxies (M51; \citealt{hughes13b}; \citealt{colombo14a}) or in the Orion-Monoceros dataset itself, this approach fails to recognize objects on the physical scale of galactic GMCs. In the latter case, particularly, most of the clouds of the complex are encompassed by the same island. Indeed, the ``island'' method is generally not sufficient to identify clouds. Therefore, the islands are subsequently divided into ``clouds'', or independent local maxima within the islands, through a watershed algorithm. Only pixels that can be uniquely associated to a given local maximum will constitute the final ``cloud'', while shared pixels are discarded and considered as not being part of any structure (forming the so-called ``watershed''). This approach has been very successful for many (mostly extragalactic) applications (e.g. \citealt{bolatto08}, \citealt{wong11}, \citealt{gratier12}, \citealt{rebolledo12}, \citealt{colombo14a}). Here the CPROPS decomposition is performed using physically-motivated priors\footnote{From \cite{rl06} table 2, local maxima must have a distance of at least 15\,pc between each other, a velocity separation of 2\,km\,s$^{-1}$ and a significance of at least 1\,K.} that should provide objects closer to what is thought to be a GMC. Nevertheless in the case of Orion-Monoceros the emission appears too over-divided and the identified objects seem to be more comparable to dense clumps within the clouds rather than actual GMCs. 

The CPROPS package implements also the original version of CLUMPFIND by \cite{williams94}. The CLUMPFIND algorithm uses a friends-of-friends procedure to decompose the emission within a single cloud or cloud complex. This algorithm contours the data into a finite number of intensity steps, assuming a one-to-one relation between peaks in the intensity profile and clumps. At the ``blending level'' between two or more clumps, the flux is equally distributed between the clumps. Unlike the CPROPS decomposition, CLUMPFIND conserves flux, so that all the flux within the data is assigned to individual clumps. Being designed to identify clumps within clouds, CLUMPFIND is less suitable for the GMC decomposition.  When applied to rich, structured data, CLUMPFIND tends to provide ``patchwork'' segmentations that have little resemblance to physical structures.    

\begin{figure*}
\centering
\includegraphics[width=1\textwidth]{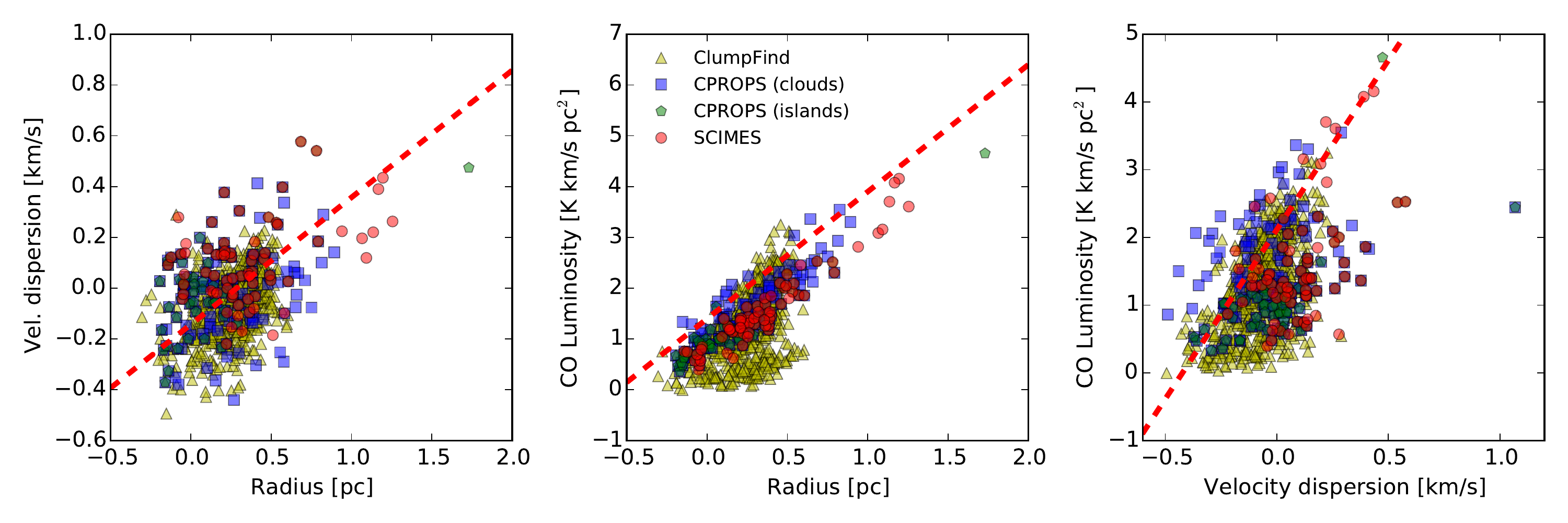}
\caption{Scaling relations comparison between the different segmentation algorithms. Red dashed lines indicate the fit obtained from Galactic clouds  (Solomon et al. 1987): ($\sigma_{v}$/[km/s]) = 0.72 ($R/$[pc])$^{0.5}$ for the left panel, ($L_{\mathrm{CO}}$/[K km/s pc$^{2}$]) = 25 ($R/$[pc])$^{2.5}$ for the central panel, ($L_{\mathrm{CO}}$/[K km/s pc$^{2}$]) = 130 ($\sigma_{v}/$[km/s])$^{5}$ for the right panel.}
\label{F:orion_algo_comp_props}
\end{figure*}

%---------------------------------------------------------------------
\subsection{Appearance of scaling relations and mass spectra using different segmentation methods}
\label{SS:algo_comp_props} 
%---------------------------------------------------------------------
We now compare the properties of the objects obtained with different segmentation methods\footnote{For this analysis we do not consider the results given by the dendrogram itself.}, and in particular to discuss the appearance of the derived scaling relations and mass spectra. Since early studies of GMC population (e.g., \citealt{solomon87}), the scaling relations (also called ``Larson's laws'' from \citealt{larson81}) and mass spectra have been standard tools to investigate the physical state of these objects and also to diagnose their formation and evolution \citep[e.g.,][]{gratier12,colombo14a}. Nevertheless, several studies \citep[in particular][]{hughes13b} have demonstrated that, in complex environments, their appearance is largely biased by instrumental sensitivity, resolution, and the method used to decompose the clouds. Therefore, it is necessary to understand how the properties of the objects segmented by \texttt{SCIMES} compare with other methods. For a fair comparision that depends only on the identification methods, we calculate the physical properties of the clouds identified by each algorithm using the same moment method, as implemented in the \verb"cloudalyze" procedure of CPROPS (\citealt{rl06}).

\begin{figure}
\centering
\includegraphics[width=0.45\textwidth]{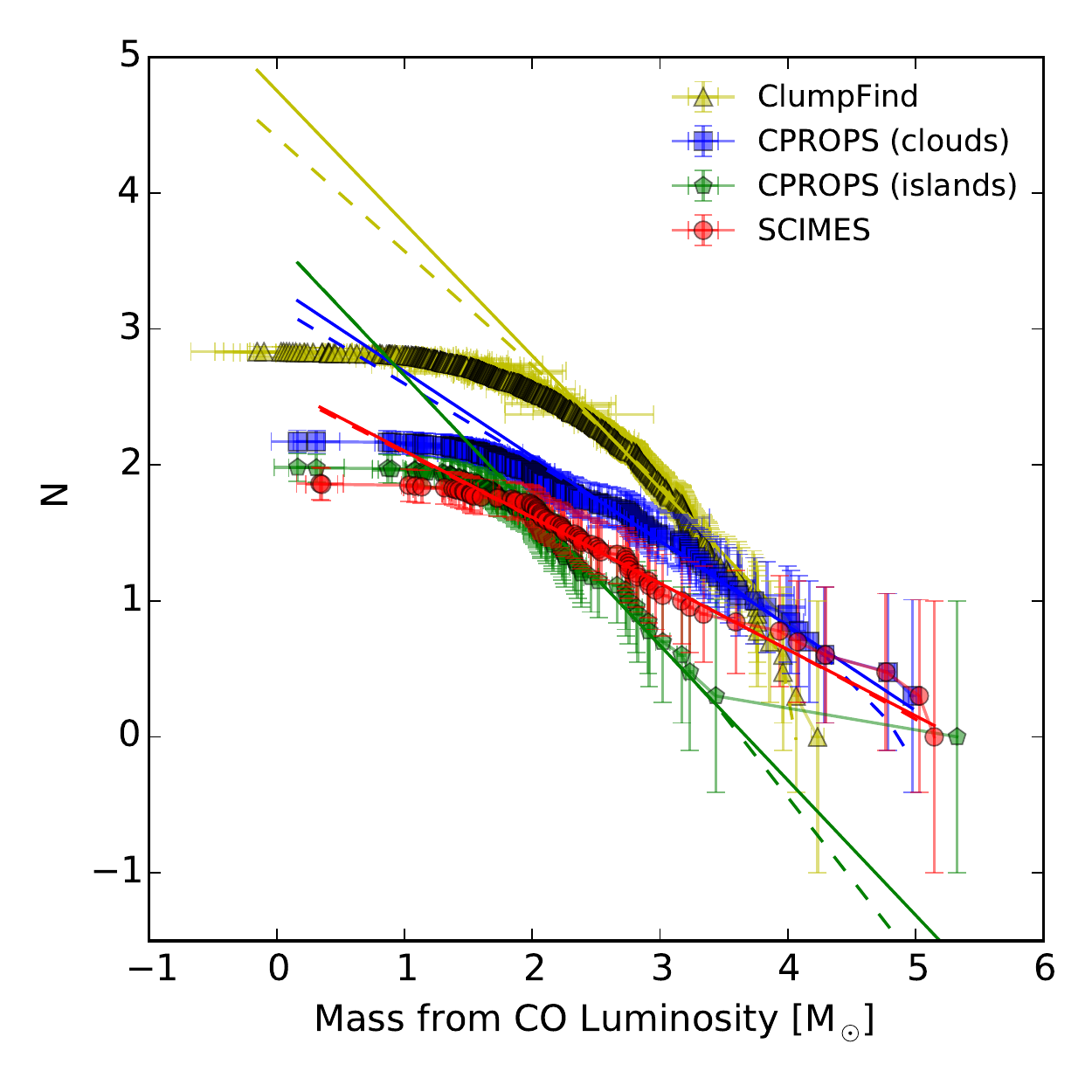}
\caption{Mass spectra comparison.  The shapes of the mass spectra are determined by the nature of the algorithm.  CLUMPFIND tends to find small clumps and the influence of the mode choice CPROPS shows up clearly in the shapes identified.  \texttt{SCIMES} finds many of the canonical clouds in the area, but the mass distribution is likely incomplete at the low-mass end. The full line indicates the power law fit from equation~\ref{E:pl}, while the dashed line its truncated version of equation~\ref{E:tpl}.}
\label{F:orion_algo_comp_mspecs}
\end{figure}

The basic properties of the objects that we plot are combined into scaling relations (figure~\ref{F:orion_algo_comp_props}) and mass spectra (figure~\ref{F:orion_algo_comp_mspecs}). We do not make any attempt to correct those properties for the survey biases (as described in \citealt{rl06}; R08) since we are interested in the properties provided by different segmentation approaches rather than comparing datasets with different observational biases.  Generally the scaling relations show significant scatter. In particular, none of the analysis methods yield compelling evidence for a tight size-line width or luminosity-line width correlation. Linear relationships between virial mass and CO luminosity, and luminosity mass and radius are evident from \texttt{SCIMES} and CPROPS in ``cloud'' mode. CPROPS in ``island'' mode is, instead, dominated by a single large object (see figure~\ref{F:orion_algo_comp_asgn}) that encompass most of the structures in the complex, the properties of which can be seen as outliers for this dataset. For \texttt{SCIMES} and CPROPS decompositions, the scatter in the scaling relations is partially due to small, isolated emission features of uncertain nature that typically have the size of the resolution element (but see next Section). Looking at the dendrograms in figure~\ref{F:orion_dendro}, those objects emerge mostly in the trunk and are not related to the main emission branch. Using \texttt{SCIMES}, they can be easily eliminated (a posteriori). To some extent, this cleaning operation is also possible after the CPROPS segmentation, if it is performed using physical motivated priors. All structures decomposed by CLUMPFIND have properties very close to the resolution element. CLUMPFIND is, indeed, the algorithm most influenced by the survey designs of the ones considered here. At this resolution ($\sim1$\,pc), however, CLUMPFIND divides the emission into objects with the size of the so-called ``clumps'' generally considered as the born places of stellar clusters (\citealt{williams94}). 

The CLUMPFIND mass spectrum presented in figure~\ref{F:orion_algo_comp_mspecs} can be regarded as a genuine ``clump'' spectrum. The \texttt{SCIMES} spectrum instead characterizes isolated and independent entities closer to the classical definition of molecular clouds. Nevertheless, this cannot be assumed as a representative GMC spectrum since, in this dataset, only few objects on the characteristic scale of a GMC are present, resulting in a undersampled spectrum at the lower end. CPROPS (in ``cloud'' mode) provides a collection of objects halfway between clumps and molecular clouds, close to the values of the physical priors imposed. Its spectrum might be representative of compact objects on a scale up to $\sim10$\,pc. The CPROPS spectrum for ``islands'' is clearly biased by the presence of the central object. However, it can be used to trace the the mass contribution of small isolated objects close to the size of the resolution element. 

To formally test these trends, we fit the mass spectra using both a power-law:

\begin{equation}\label{E:pl}
N(M' > M) = \bigg(\frac{M}{M_0}\bigg)^{\gamma+1},
\end{equation}

\noindent and its truncated version:

\begin{equation}\label{E:tpl}
N(M' > M) = N_0\bigg[\bigg(\frac{M}{M_0}\bigg)^{\gamma+1} - 1\bigg].
\end{equation}

\noindent adapted for cumulative mass distributions (\citealt{rosolowsky05}, \citealt{wong11}). In these equations $M_0$ represents the maximum mass of the sample. In the truncated version of the model, $M_0$ indicates also the mass where the spectrum rolls of, since $N(M\geq M_0)=0$. In this case, $N_0$ is the number of clouds more massive than $2^{1/(\gamma+1)}M_0$. Equations~\ref{E:pl}-\ref{E:tpl} are integrals of differential cloud mass distribution, $dN/dM\propto M^\gamma$. The index $\gamma$ is considered as an indicator for the general mass of the molecular cloud ensemble: $\gamma>-2$ means that most of the molecular gas of the complex is enclosed into massive GMC, while $\gamma<-2$ indicates that small entities dominate the molecular mass budget. We fit the two models using the Orthogonal Distance Regression\footnote{http://docs.scipy.org/doc/scipy/reference/odr.html} implemented in \verb"scipy", that accounts for uncertainties in both variables. Errors on the mass from CO luminosity are generated through CPROPS \verb"cloudalyze" procedure using 100 bootstrap iteration. The cumulative number instead as an error characterized by a counting error given by $\sqrt{N}$. Given the survey designs, (see Section~\ref{S:test}), we fit the models above a mass of 100\,M$_{\odot}$. Results of the fit are shown in figure~\ref{F:orion_algo_comp_mspecs}. Both models give almost indistinguishable indexes for each related method segmentations. Nevertheless, for CLUMPFIND $\gamma\sim-2.5$, and for CPROPS (in ``island'' mode) $\gamma\sim-2.1$. Instead \texttt{SCIMES} shows $\gamma\sim-1.5$, similar to CPROPS in ``cloud'' mode where $\gamma\sim-1.6$. Therefore, according to the result of CLUMPFIND, one would tend to deduce that the molecular mass of the Orion-Monoceros complex is mostly enclosed into small objects. The opposite interpretation, however, is suggested by the fits of \texttt{SCIMES} and CPROPS in ``cloud'' mode. CPROPS in ``island'' mode provides results exactly on the border, indeed most of the entities that compose its mass spectrum are small objects, while the fit is influenced by the large structure that encompasses almost all molecular gas of the survey. In conclusion, different algorithms might give largely different results, and they have to be applied with care according to the characteristic of the data under analysis. 
 
%=====================================================================
\section{Hierarchical scaling relations}
\label{S:scal_rel} 
%=====================================================================

In the previous Section we showed that every segmentation method (including \texttt{SCIMES}) introduces an amount of scatter in the scaling relations between the properties of the objects. This is particularly true for the size-line width relation, since these two quantities are not covariant. Nevertheless, being based on the dendrogram framework, \texttt{SCIMES} is a multi-scale decomposition method that explicitly takes the hierarchical nature of the ISM into account. Therefore, we combine this two approaches to investigate the appearance of Larson's laws within the hierarchy of the clouds.  

\begin{table*}
\begin{tabular}{ccccccc}
\hline
Object & \multicolumn{2}{c}{$\sigma_{v}=(a_{1}\pm\delta a_{1})R^{(b_{1}\pm\delta b{1})}$} & \multicolumn{2}{c}{$L_{CO}=(a_{2}\pm\delta a_{2})R^{(b_{2}\pm\delta b_{2})}$} & \multicolumn{2}{c}{$L_{CO}=(a_{3}\pm\delta a_{3})\sigma_{v}^{(b_{3}\pm\delta b_{3})}$} \\
\hline
 & $(a_{1}\pm\delta a_{1})$ & $(b_{1}\pm\delta b_{1})$ & $(a_{2}\pm\delta a_{2})$ & $(b_{2}\pm\delta b_{2})$ & $(a_{3}\pm\delta a_{3})$ & $(b_{3}\pm\delta b_{3})$ \\
\hline
Orion A & 0.40$\pm$0.01 & 0.50$\pm$0.04 & 14.33$\pm$0.73 & 2.06$\pm$0.10 & 575.63$\pm$25.6 & 4.02$\pm$0.21 \\
Orion B & 0.66$\pm$0.01 & 0.46$\pm$0.02 & 21.04$\pm$0.59 & 2.47$\pm$0.03 & 169.49$\pm$9.71 & 4.58$\pm$0.19 \\
Monoceros & 0.61$\pm$0.01 & 0.35$\pm$0.01 & 12.89$\pm$0.38 & 2.07$\pm$0.04 & 204.11$\pm$11.33 & 5.60$\pm$0.45 \\
North. Filament & 0.36$\pm$0.01 & 0.47$\pm$0.03 & 11.76$\pm$0.64 & 2.31$\pm$0.05 & 1422.74$\pm$96.60 & 3.88$\pm$0.24 \\
NGC2149 & 0.39$\pm$0.01 & 0.69$\pm$0.01 & 18.46$\pm$0.55 & 2.50$\pm$0.03 & 558.40$\pm$16.96 & 3.52$\pm$0.10 \\
Crossbones & 0.48$\pm$0.01 & 0.41$\pm$0.03 & 12.23$\pm$0.74 & 1.95$\pm$0.08 & 357.54$\pm$39.58 & 4.59$\pm$0.57 \\
Scissors & 1.91$\pm$0.04 & 1.17$\pm$0.07 & 16.26$\pm$0.83 & 2.33$\pm$0.17 & 4.50$\pm$0.32 & 1.96$\pm$0.30 \\
Stain & 0.54$\pm$0.02 & 0.51$\pm$0.06 & 7.06$\pm$0.32 & 2.30$\pm$0.08 & 80.57$\pm$5.68 & 3.85$\pm$0.40 \\
\hline
Milky Way & 0.72 & 0.50 & 25.00 & 2.50 & 130.00 & 5.00 \\
\hline
\end{tabular}
\caption{Best-fitting parameters for the hierarchical scaling relations of figure~\ref{F:orion_showers} (left column) for the larger objects of the Orion-Monoceros complex. For comparison, the last line of the Table summarizes the best fitting parameters of Milky Way clouds (Solomon et al. 1987).}
\label{T:orion_shower_fits}
\end{table*}

\begin{figure*}
\centering
\includegraphics[width=0.75\textwidth]{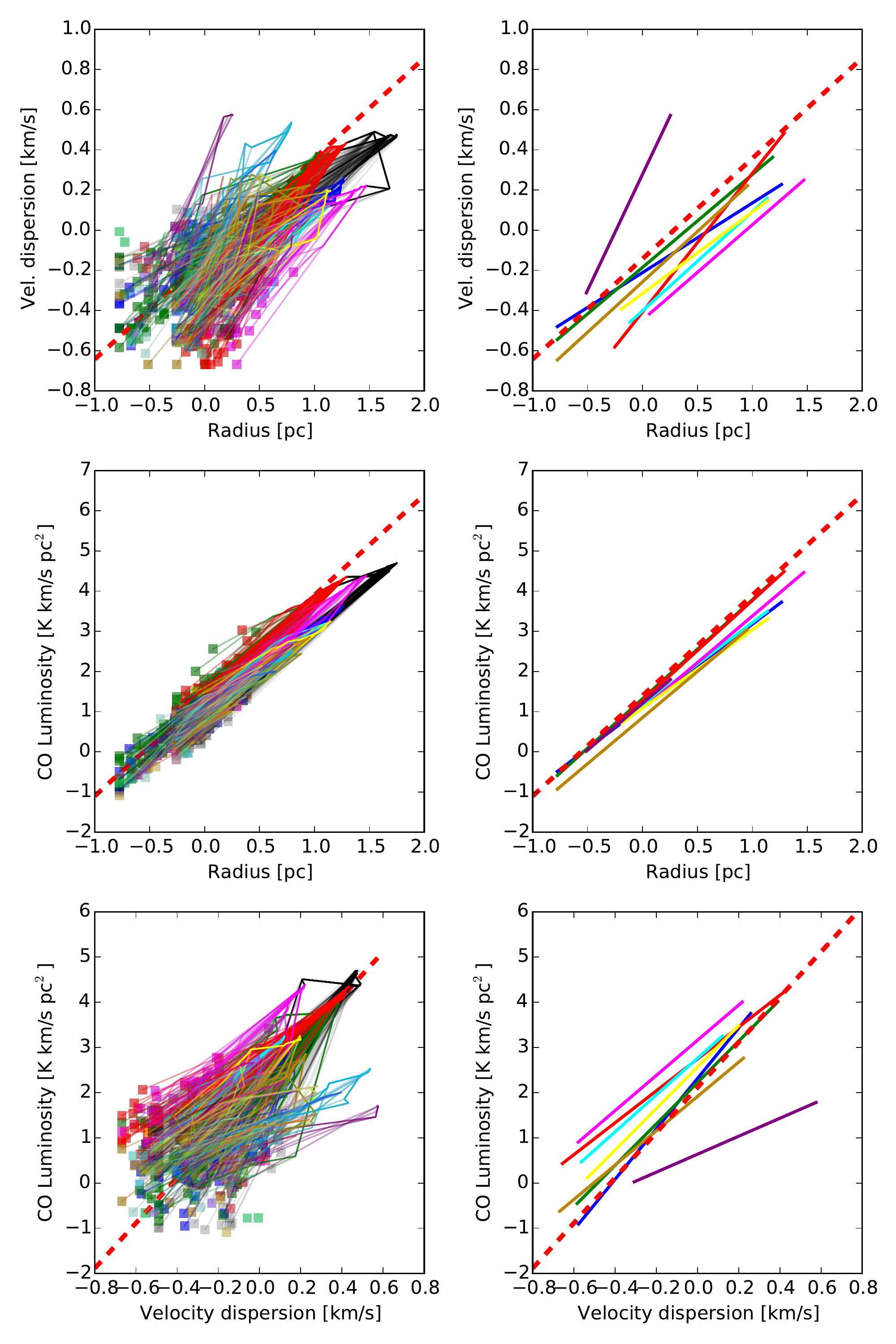}
\caption{The hierarchical scaling relations within the substructures of the Orion-Monoceros complex (left column). The straight lines between the different points of the plot represent the hierarchical relations between the various substructures of the main complex. Structures belonging to a particular cluster identified by SCIMES are indicated with the same colors as figure~\ref{F:orion_dendro} and figure~\ref{F:orion_cont}. The trunk of the dendrogram is drawn in black. In the right column, the BCES fits of the structures associated with the main objects of the complex (Orion A, Orion B, Monoceros, the Northern Filament, NGC2149, the Scissor, and Stain) are shown next the corresponding left panel. Red dashed lines indicate the fit obtained from Galactic clouds: ($\sigma_{v}$/[km/s]) = 0.72 ($R/$[pc])$^{0.5}$ for the upper panels, ($L_{\mathrm{CO}}$/[K km/s pc$^{2}$]) = 25 ($R/$[pc])$^{2.5}$ for the middle panels, ($L_{\mathrm{CO}}$/[K km/s pc$^{2}$]) = 130 ($\sigma_{v}/$[km/s])$^{5}$ for the lower panels (Solomon et al. 1987).}
\label{F:orion_showers}
\end{figure*}

Figure~\ref{F:orion_showers} (left column) recasts the relationships between cloud properties in terms of different objects within the hierarchies.  We refer to these plots as {\em shower plots}, based on their resemblance to cosmic ray showers. Each straight line in a shower indicates the hierarchical connection between two sub-structures of the dendrogram. A similar study was proposed by R08 and used by \citet{kauffmann10a} to explore the size-mass relation between several Galactic clouds. From the plots, we first note that different clusters originate from different regions of the dendrogram within the parameter space defined by a given relation. Moreover, the substructure properties within the different clusters align on well defined tracks with significantly lower scatter with respect to the cluster-to-cluster relations (see figure~\ref{F:orion_algo_comp_props}). To quantify this observation,  we calculate the Spearman rank correlation coefficient, $r_{sp}$. This coefficient evaluates how well a relation between two variables can be described by a monotonic function. If repeated values are not present within the data, $\pm1$ indicates monotonically increasing (decreasing) behaviors between the variables. As in \cite{hughes13b} and \cite{colombo14a}, we consider $r_{sp}>0.8$ as an indicator of strongly correlated properties, $0.5<r_{sp}\leq0.8$ for moderate correlated properties, and $r_{sp}\leq0.5$ for poorly correlated properties. Table~\ref{T:orion_rsp} shows that for all relations examined here, $r_{sp}$ is  high between the substructure properties of each cluster.  The cluster-to-cluster Spearman ranks are only this high ($r_{sp}>0.8$) for the relationships involving intrinsically correlated quantities (i.e. the size-luminosity and the luminosity-virial mass relations).  The coefficient is low ($r_{sp}\sim0.3$) for the remaining relationships. This large scatter, particularly in the cloud-to-cloud size-line width relationship, has been noticed in various extragalactic studies of GMCs (e.g., \citealt{hughes10}, \citealt{wong11}, \citealt{gratier12}, \citealt{hughes13b}, \citealt{colombo14a}). The coefficients related to separate clusters are always slightly higher than the ranks given by the dendrogram structure containing most of the objects of the complex. 

To test whether those trends are real and not imposed by the decomposition method, we generate several fake data cubes, setting a random power spectrum of the brightness distribution in the Fourier space, and adding a quantity of random Gaussian noise with $\sigma$rms = 0.3\,K. Then we run \texttt{SCIMES} calculating the properties of the individual clusters identified and of their substructures. We find that the monotonicity of the fake showers is high showing $r_{sp}$ similar to the ones observed for the Orion-Monoceros data. Instead, the Spearman rank within clusters in the fake data is always between 2-3 times higher than the one measured here. The size-line width relation scatter between clusters observed in the complex appears to be a real feature of the data rather than a decomposition artifact.

Thus, \texttt{SCIMES} identified objects that are not only mathematically but also physically distinct through the similarity criteria we set.  Furthermore, Larson's laws look more compelling when analyzed within the hierarchy of the clouds. It is worth noting also that the larger clusters of the complex have a well-resolved and non-trivial inner structure. Therefore the relations we observe do not arise from viewing a monolithic object at different levels of the hierarchy.

Given the monotonic relation between the structure properties within the same showers (Table~\ref{T:orion_rsp}), we fit ``Larson's Law'' relations to the individual showers to  facilitate comparison with the clouds in the Galaxy (\citealt{solomon87}).  We use the python implementation\footnote{https://github.com/rsnemmen/BCES/blob/master/bces.py} of the BCES method described in \cite{akritas96}. This method takes into account the intrinsic scatter of data and the measurement errors in both variables. For simplicity, we consider the uncertainties to be uncorrelated, even if some properties should have significant covariance. We generate errors for each property of the cataloged dendrogram structures using a bootstrapping method similar to the one described in Section 2.5 of \cite{rl06}. We use 1000 bootstrap iterations.  Figure~\ref{F:orion_showers} (right column) shows the fits of the showers from the larger objects with non-trivial hierarchies, while the result of the fit are reported in Table~\ref{T:orion_shower_fits}.  Overall, the showers show remarkably similar fitting parameters to each other. Larger differences are distinguishable mostly within the amplitude of the relations rather than between the slopes. In particular, this is the case for the structures corresponding to the Scissors, which occupies its own position within the parameter space in most of the relationships.  This gas structure can be seen as an outlier of the complex. Indeed, the Scissors has been interpreted as a superposition of two distinct objects along the line-of-sight, given its complicate kinematics (\citealt{wilson05}).  

The fit we performed within the showers has a slope almost indistinguishable from the size-line width relation of \cite{solomon87} or \cite{heyer_brunt04}, that might indicate that Orion-Monoceros objects are dominated by the same kind of turbulence of the objects observed in those studies (such as the Burgers turbulence in supersonic conditions; \citealt{passot88}). However we also notice a large scatter on the cluster-to-cluster first ``Larson's law'' (very low $r_{sp}$ value, see Table~\ref{T:orion_rsp}), i.e. on a scale corresponding to $\sim5-10$\,pc. In general the scatter is always present in the relationships involving a direct measure of the velocity dispersion (like the luminosity-velocity dispersion relation). This might reflect the fact that, at least for the Orion-Monoceros objects, the line width is tracing internal differences between the various clouds. Those differences might be imposed by distinct levels of stellar feedback that vary the energy injected in the gas, and modifying the amplitude of the turbulence fluctuations from cluster to cluster.

\begin{table}
\centering
\begin{tabular}{cccc}
\hline
Object & $r_{sp}^{1}$ & $r_{sp}^{2}$ & $r_{sp}^{3}$ \\
\hline
Orion A & 0.84 & 0.97 & 0.91 \\
Orion B & 0.84 & 0.96 & 0.82 \\
North. Filament & 0.87 & 0.98 & 0.90 \\
Monoceros & 0.87 & 0.97 & 0.88 \\
NGC2149 & 0.90 & 0.97 & 0.90 \\
Crossbones & 0.80 & 0.92 & 0.82 \\
Scissors & 0.96 & 0.93 & 0.88 \\
Stain & 0.78 & 0.96 & 0.81 \\
\hline
Clusters & \textbf{0.26} & 0.93 & \textbf{0.35} \\
Trunk & 0.82 & 0.97 & 0.82 \\
\hline
\end{tabular}
\caption{Spearman's rank correlation coefficients for the substructures within the most notable objects of the complex. $r_{sp}^{1}$ refers to $\sigma_{v}=a_{1}R^{b_{1}}$, $r_{sp}^{2}$ to $L_{CO}=a_{2}R^{b_{2}}$, $r_{sp}^{3}$ to the $L_{CO}=a_{3}\sigma_{v}^{b_{3}}$ relationship. The line of the Table, labeled ``Clusters'' indicate the cluster-to-cluster Spearman's rank for the same relationships, while the last line (``Trunk'') denotes the Spearman rank of the dendrogram structure containing most of the clouds of the complex (drawn with black lines in figure~\ref{F:orion_showers}).}
\label{T:orion_rsp}
\end{table}

%=====================================================================
\section{Discussion and Outlook}
\label{S:disc} 
%=====================================================================

In this paper we present \texttt{SCIMES} (Spectral Clustering for Interstellar Molecular Emission Segmentation), which introduces a novel approach to segment the molecular medium based on dendrograms, graph theory, and clustering. We also introduced many new concepts for the study of the ISM connected with the cloud decomposition and the multi-scale analysis of the molecular gas emission. Here we  discuss their scientific implications and the unique possibilities offered by the method. 

%---------------------------------------------------------------------
\subsection{Unique features and limitations of the algorithm}
\label{SS:disc_feats}
%---------------------------------------------------------------------

One of the main features of \texttt{SCIMES} is to provide objects with sizes significantly larger than the resolution element and not simply determined by the sensitivity of the data. This distinguishes our method from other algorithms for ISM segmentation that are very much constrained by the size of the beam and channel width of the data set. Indeed, clustering means  grouping together entities that are similar and separating them from others which show a lower level of similarity. The clusters identified by \texttt{SCIMES} will be, in general, always larger than the resolution element. In a multi-scale ISM, these objects can possess complex morphologies. In the Orion-Monoceros dataset, we found compact, round clouds as well as concave and elongated shapes. This unique feature of the algorithm allow for a more morphologically-oriented study of the molecular medium.  Nevertheless, for most extragalactic observations, the beam size is roughly the characteristic size of the GMCs.  Then, the leaves of the dendrogram already represent single clouds. In this case, the clusters would be collections of GMCs that might be useful for some applications and studies, but not for a GMC catalog.  Other algorithms would be more appropriate for identifying GMCs in that use case.  

\texttt{SCIMES} identifies structures that can be considered as single entities since the dendrogram leaves that compose these structures can be grouped together according to cluster theory.  \texttt{SCIMES} labels some dendrogram structures as independent from others relative to the similarity criteria chosen. Those structure are, therefore, already present and cataloged by the dendrogram, and \texttt{SCIMES} cannot find other objects that are not considered by the dendrogram. Since the dendrogram is constructed considering the pixel neighborhood according to the dimension of the dataset, the structures identified by \texttt{SCIMES} in a PPV cube are velocity-connected. However, velocity-connected objects (especially in the Galaxy) are not always at the same distance as in the case of the Orion-Monoceros complex (but see Appendix~\ref{A:nodist}). If distances are known they can be provided as input to define the similarity matrices. The final segmentation would be more physically-oriented rather than data-oriented as for the Orion-Monoceros segmentation proposed in Section~\ref{S:test}. 

The flexibility of the dendrogram to operate on multidimensional datasets makes \texttt{SCIMES} already applicable for \emph{position-position} images (converting, for example, the ``volume'' criteria into an ``area'' criteria) or \emph{position-position-position} simulated cube (e.g. Duarte-Cabral et al. in prep.). Moreover, future developments for more simulation-oriented applications in the PPPV or PPPVVV domains are possible.

The dendrogram-basis of \texttt{SCIMES} offers other advantages. The hierarchical structure is one of the main characteristics of the molecular medium that reflects its turbulent nature. Within the identified objects, this structure is readily available after the dendrogram creation. This might allow, for example, to study how turbulent energy is transferred and dissipated from the size of the clouds to smaller, inner, bound structures \citep{rosolowsky08}. 

While cluster analysis is a novel approach in astronomy, \texttt{SCIMES} has been developed with standard clustering algorithms adapted from other disciplines to suit our needs.  The data mining and machine learning literatures are rich in alternatives to spectral clustering (e.g., \citealt{jain99}), k-means (e.g., \citealt{hamerly}), and the silhouette index (e.g. Desgraupes 2013). Our method might be improved or complemented in a number of different ways.

%---------------------------------------------------------------------
\subsection{The physical meaning of the similarity matrix and the scaling parameter}
\label{SS:disc_affmats}
%---------------------------------------------------------------------

\texttt{SCIMES} starts by abstracting the dendrogram obtained from a star-forming complex into a graph, where the leaves of the dendrogram are seen as vertices of the graph and the hierarchical level where two leaves merge defines the connection (edge) between the nodes. A key step that strongly determines the final segmentation results is the choice of the similarity criteria or equivalently the weights of the graph edges. The current implementation of \texttt{SCIMES} makes use of two criteria, based on the PPV volume of the smallest isosurface containing the two leaves under consideration and the amount of flux (or CO luminosity, when distance information is available) within it. Those criteria are readily available from the dendrogram itself and appear to provide good results. Moreover, being cumulative properties, ``volume'' and ``luminosity'' provide well-behaved, block diagonal similarity matrices. In particular, within the volume-associated matrix the blocks corresponding to the classic clouds of the Orion-Monoceros complex stand out clearly and \texttt{SCIMES} has no difficulty in identifying them. 

The spectral clustering also requires insignificant cluster affinities to be rescaled out using, for instance, a Gaussian kernel (or its alternative version used in the paper). The Gaussian kernel has a free ``scaling parameter'' to be set for this operation. We selected an optimal value for this parameter with an initial guess from the similarity matrix.  In the case of the Orion-Monoceros complex, a volume of $~\sim 8000$\,pc$^{2}$\,km/s corresponds to an effective radius $R_{eff}\sim30$\,pc and a velocity dispersion $\sigma_{v}\sim3$\,km/s. Those quantities are very close to the usually cited characteristic scales of the GMCs (e.g \citealt{blitz07}) and \texttt{SCIMES} finds them through a pure data-driven analysis. Given the clearly defined appearance of the affinity matrix, it seems that the objects in this star forming complex have a well-established maximum size and velocity dispersion.  If a larger scale in the molecular hierarchy existed, the data of sufficiently wide area that \texttt{SCIMES} would be able to find it. In the same way, from the luminosity criterion, \texttt{SCIMES} selects a scaling parameter equal to 28128\,K\,km/s\,pc$^{2}$, equivalent, by assuming a Galactic $\alpha_{CO}=4.4$\,M$_{\odot}$/(K\,km/s\,pc$^{2}$) (e.g. \citealt{strong_mattox96}), to $\sim1.2\times10^{5}$\,M$_{\odot}$, again similar to the average mass assumed for the GMCs in the Milky Way. \texttt{SCIMES} finds those parameters, as well as the initial guess for the number of clusters, automatically via a direct analyses of the affinity matrix. These variables can be also imposed by the user if necessary. 

For the Orion-Monoceros complex dataset we analyzed here, the blocks of the similarity matrix associated with the luminosity criterion are not as well defined as in the volume matrix. For this dataset, the luminosity is not a clustering criterion as good as the volume. Additionally, the silhouette calculated for the best clustering configuration through the luminosity criterion is not as high as the one obtained from the volume criterion alone (see Section~\ref{SS:test_oriondata}). The silhouette profile shown in Figure~\ref{F:orion_siltest} for the luminosity presents ambiguities that are not observed in the profile of the volume criterion. Taken together, these evidences might indicate that for the particular tracers and resolution used to image the considered star forming region, the emission is better segmented through its morphological features, rather than the emission. For the given observation and scale, the relevant structures tend to have similar volumes rather than luminosities. 

In this aspect, the differences in \texttt{SCIMES} performance between the ``luminosity'', and ``volume'' similarity criteria might be generally connected with the question of \emph{which is the ``right'' segmentation criterion for the clouds?}  This question can be more generally posed as \emph{what is the physical mechanism responsible for the clumpiness of the molecular ISM?} or \emph{are molecular clouds real, independent entities, and an important scale for the star formation process?} Even so, the different criteria might produce segmentations that mimic a by-eye approach without being linked to a particular physical origin.  Nevertheless, the most innovative feature of \texttt{SCIMES} is the possibility to expand the friends-of-friends segmentation concept, where the ``friendship'' is set not simply by the value of neighboring pixels, but by real physical properties of the ISM.  Indeed, the main strength of the spectral clustering is to shift every property that can be seen as a similarity into an Euclidean space where clustering features are enhanced, with basically no restrictions.  Moreover, the isosurfaces associated with the graph edges are well-defined three-dimensional structures that posses their own physical properties.  Those properties can be applied to construct customized similarity matrices to be used to generate segmentation based on the physics one wants to explore. For example, similarity matrices can be obtained considering the amount of star formation within the isosurface, the abundance of a particular chemical tracer, the level of dust extinction, or kinematic properties.  The products of such segmentation would be gas ``clusters'' having a characteristic maximum property defined by the similarity criteria utilized. Moreover, through the matrix aggregation, various segmentations can be obtained aggregating several criteria together. The dominant one(s) (as in the case of ``volume'' versus ``luminosity'') would leverage lower similarities among the others. 

%---------------------------------------------------------------------
\subsection{Toward an unified definition for the molecular gas structures: Molecular Gas Clusters}
\label{SS:disc_mgc}
%---------------------------------------------------------------------

According to their properties, the objects identified by \texttt{SCIMES} in the Orion-Monoceros complex match with the general definitions of ``Giant Molecular Cloud'' present in the literature. Indeed, from the theoretical point of view, GMCs can be seen as the largest star formation-coherent regions \citep{kruijssen_longmore14}, of which the size is expected to be characterized by the \citet{toomre64} wavelength. For a Toomre stability parameter $Q$ close to unity, this length-scale is roughly equal to the scale height of the gaseous disk (\citealt{krumholz_mckee05}). For observations, however, the working definition of \citealt{williams00} is usually adopted: GMCs are considered as appropriate sites of star formation, formed mostly from molecular gas, having masses $\geq 10^{4}$\,M$_{\odot}$, and sizes $\sim20-50$\,pc, whose properties are noticeably different from the ambient medium (\citealt{kennicutt_evans12}). Although the latter is used to design surveys aiming for GMC studies, it is not applicable in general, and many examples of objects that deviate for one or more of the characteristics listed above are found (e.g. \citealt{blitz84}; \citealt{rosolowsky_blitz05}; \citealt{dobbs11}; \citealt{hughes13b}; \citealt{meidt13}). 

Through this paper we proposed that GMCs can also be defined using the methodologies of the cluster analysis. Conceptually, this new definition provides immediate advantages. First, it prompts GMCs with a solid mathematical formalism based on the framework of cluster analysis. Second, the act of finding clusters might help to constrain theories of GMC formation. Clustering means to group together objects or variables that share some observed qualities. Alternatively, clustering means to partition or to divide a set of objects or variables into mutually exclusive classes whose boundaries reflect differences in the observed qualities of their members. In a similar fashion, a set of gas clumps can be grouped together if they possess common properties. If this is true, it is logical to think that those clumps might also share a common origin. Various sets of common properties define a single entity (e.g. a GMC, regarding the scale under consideration), the elements of which have followed a common evolutionary path, possibly with a single outcome (e.g., a coherent star formation). If some properties are shared but not others, the clumps might originate from the same phenomena, but they have followed different evolutionary paths perhaps shaped by different environmental conditions.  Instead if no property is shared, formation and evolution might be completely separate for the group of clumps under consideration. These properties can be distinguished into location in the PPV space and physical properties of the clumps, both encoded in \texttt{SCIMES} through the dendrogram and the specific set of affinity matrices considered, respectively.

In Section~\ref{SS:test_robustpar} we found out that a resolution of 10\,pc or better is needed for the algorithm to identify GMC-like objects. In the same way, not only GMCs, but also their sub-structures such as clumps or filaments, might be associated with specific affinity matrices or clustering criteria, providing enough dynamical range to allow \texttt{SCIMES} to correctly characterize the hierarchical nature of their emission. This might help to find a common mathematical and physical definition for all molecular gas structures which could be viewed as subclasses of the more extended concept of ``Molecular Gas Clusters''. With this term we mean discrete regions of the molecular ISM that share common properties as defined by a set of similarity matrices, including possibly a common formation and/or evolutionary history.

%=====================================================================
\section{Summary}
\label{S:summary} 
%=====================================================================

We presented a generalization of the GMC segmentation problem based on graph theory and cluster analysis, to create \texttt{SCIMES} (Spectral Clustering for Interstellar Molecular Emission Segmentation).  \texttt{SCIMES} is a novel and robust approach that faithfully reproduces the work of by-eye identification of GMCs using dendrograms of emission.  Dendrograms can be seen as mathematical graphs by considering the leaves as the vertices of the graph. The edges of the graph can be weighted using the properties of the highest-level isosurface containing each pair of leaves.  Those weights are collected into similarity matrices and passed to the spectral clustering. Spectral clustering produces optimal cuts of the structure tree, which identifies the molecular clouds, while respecting the hierarchy of the dendrogram structures. We tested the method using data of the $^{12}$CO(1-0) emission from the Orion-Monoceros complex. We found that all canonical clouds of the complex (e.g., Orion A, Orion B, Monoceros, the Northern Filament, NGC2149, the Crossbones and the Scissors) are correctly recognized by the algorithm. Their properties were robust to changes in the dendrogram-generation parameters and different noise realizations.  The results are quite stable in degrading the spatial resolution by a factor 10 but performance declines for resolutions $> 10$pc.  

\texttt{SCIMES} performs best in complex environments and with high resolution data, such as those as provided by Galactic plane surveys.  This approach is thus complementary to other algorithms like CLUMPFIND and CPROPS.  When applied to well-resolved GMC data, CLUMPFIND and CPROPS (in ``cloud'' mode) tend instead to over-divide the molecular emission.  This behavior changes the shape of the mass spectra, which is more closely related to resolution-element-sized clump spectra rather than cloud spectra for high resolution surveys.  All algorithms, however, introduce a significant amount of scatter in the scaling relations between cloud properties.  We interpreted the scatter given by the properties of the objects identified by \texttt{SCIMES} as the result of real physical differences in the Orion-Monoceros clouds that leave traces in the measurement of the velocity dispersion.  The scaling relations {\em within} the hierarchies of the different objects show much tighter correlations. 

\texttt{SCIMES} finds coherent regions within data cubes. Those regions possess similar values of volume or integrated CO luminosity. The regions (clouds) decomposed by the algorithm are quite similar if they are identified by the volume, or the luminosity of both criteria aggregated. Nevertheless, the volume criterion appears to provide better clustering performance.  \texttt{SCIMES} offers also an opportunity to expand the ``friends-of-friends'' paradigm from the pixel neighborhood-based one to the physics-based one. Indeed, similarity matrices can be generated through virtually every property of the ISM, including star formation rate and chemical content.  This operation also expands the concept of ``Giant Molecular Cloud'' itself to be included in the broader class of the ``Molecular Gas Clusters''. We defined Molecular Gas Clusters as a category of discrete objects within the molecular ISM that share several common physical properties and can be segmented by a well-defined set of similarity criteria.    

The algorithm is publicly available\footnote{http://github.com/dcolombo/scimes/} and it is readily usable not only for PPV data cubes but also for the object identification in PP images and PPP simulations. In the same way, it can be tuned to recognize clumps or filaments, given enough dynamic range necessary for the correct construction of the gas emission hierarchy within the data.

%=====================================================================
% ACKNOWLEDGMENTS
%=====================================================================

\section*{Acknowledgements}
The authors thank the anonymous referee for the constructive comments that significantly help to improve the quality of the paper. The authors thank also Veselina Kalinova, Diederik Kruijssen, Abou-Moustafa Karim, Jaime Pineda, Andreas Schruba, Jens Kauffmann, Ke Wang, Volker Ossenkopf, Stephanie Walch, Friedrich Wyrowski, Arnaud Belloche, Alexander Karim, Alvaro Sanchez-Monge, Laszlo Szucs, Sharon Meidt, Andrey Shoom for the useful insight and discussions during the development of the method and the writing of the paper.  ER and DC are supported by a Discovery Grant from NSERC of Canada and research funding from the Faculty of Science at the University of Alberta. ADC acknowledges funding from the European Research Council for the FP7 ERC starting grant project LOCALSTAR. AH acknowledges support from the Centre National d'Etudes Spatiales (CNES) and funding from the Deutsche Forschungsgemeinschaft (DFG) via grants SCHI 536/5-1 and SCHI 536/7-1 as part of the priority program SPP 1573 ``ISM-SPP: Physics of the Interstellar Medium''.
%====================================================================
%=====================================================================
% REFERENCES
%==============================BIBTEX=======================================
\footnotesize{
\bibliographystyle{mn2e_new}
\bibliography{../../cold}
}

%=====================================================================
% APPENDICES
%=====================================================================
\appendix

\newpage

%=====================================================================
\section{SCIMES segmentations without distance information}
\label{A:nodist} 
\begin{figure}
\begin{center}
{\includegraphics[width=0.4\textwidth]{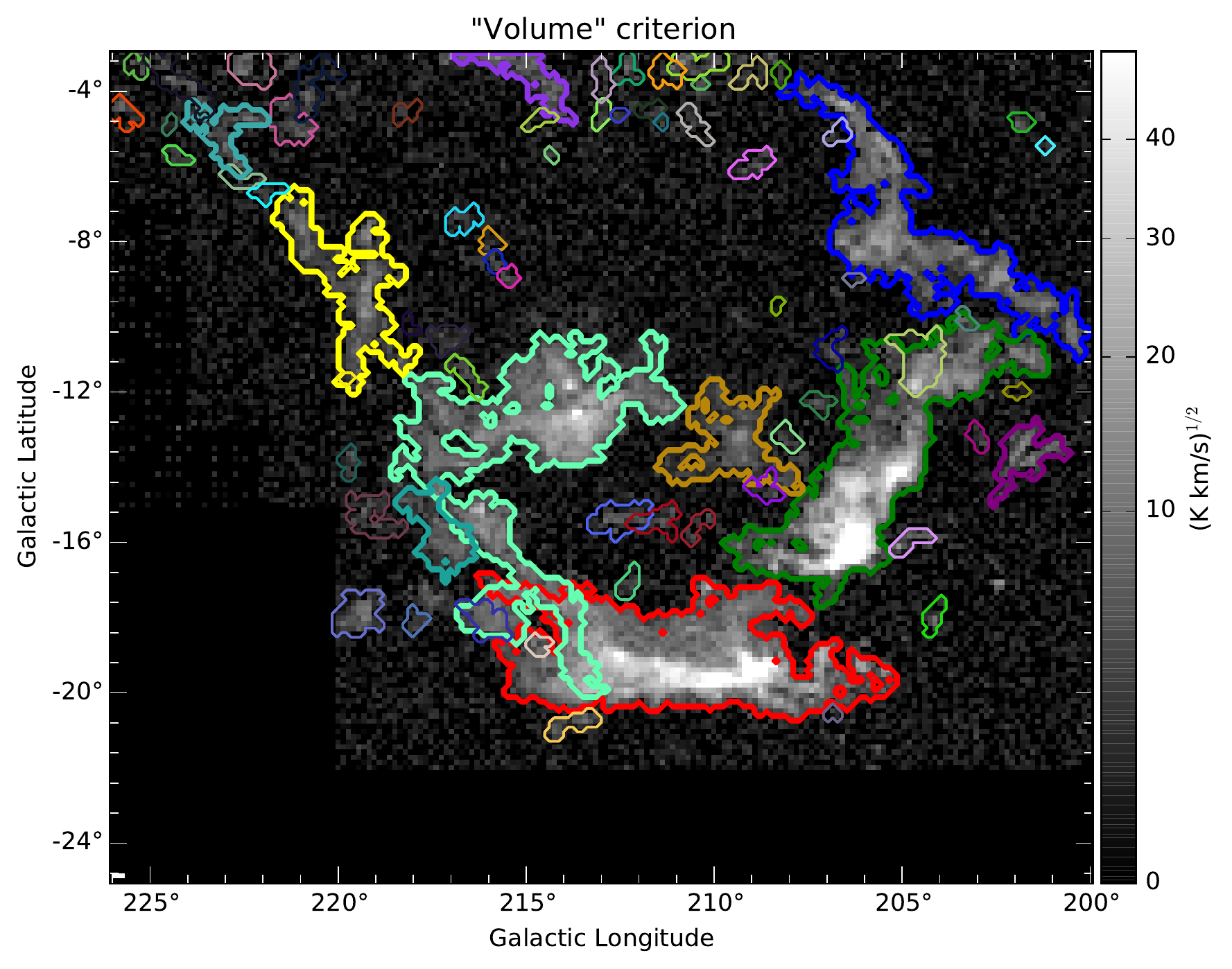}}
{\includegraphics[width=0.4\textwidth]{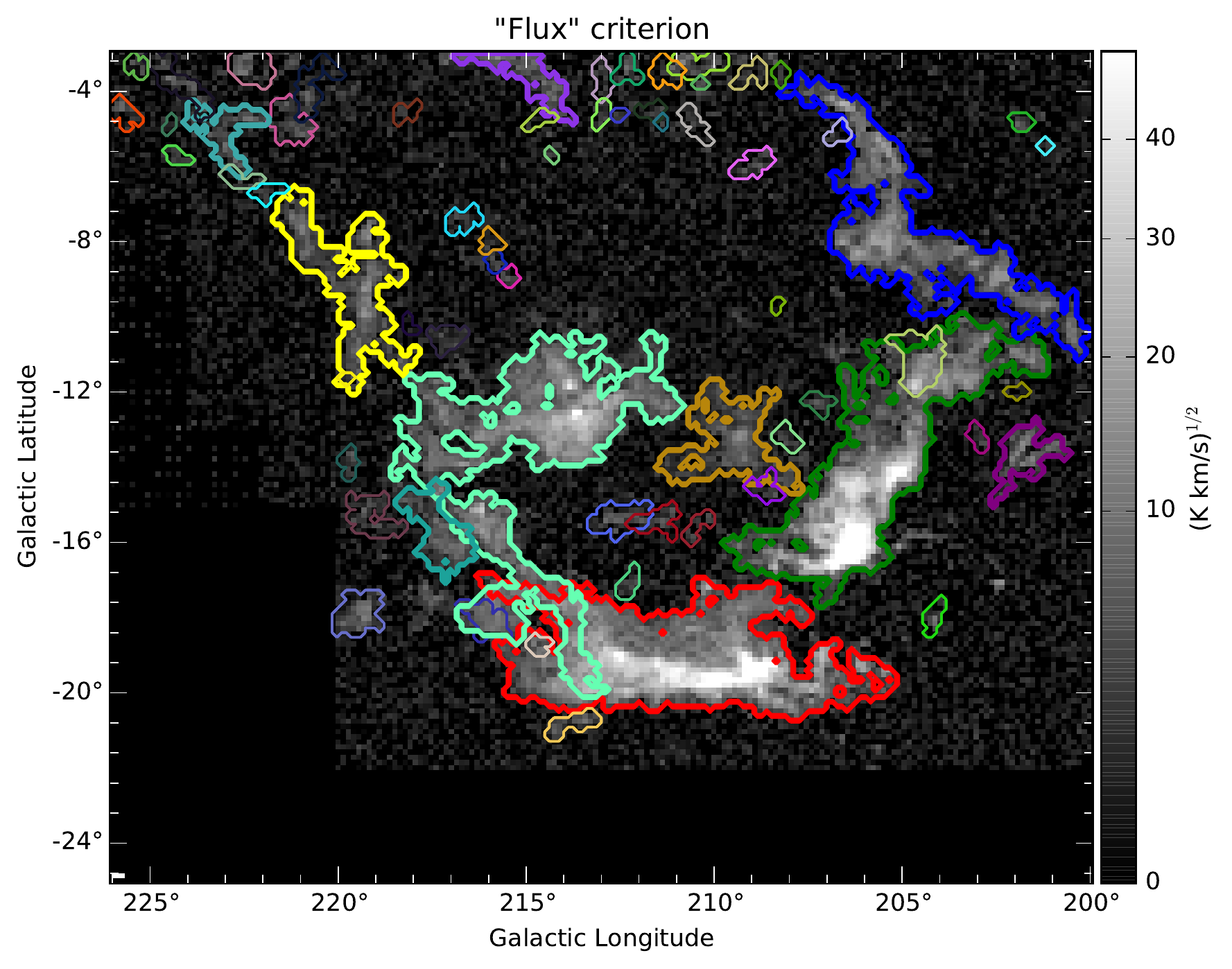}}
{\includegraphics[width=0.4\textwidth]{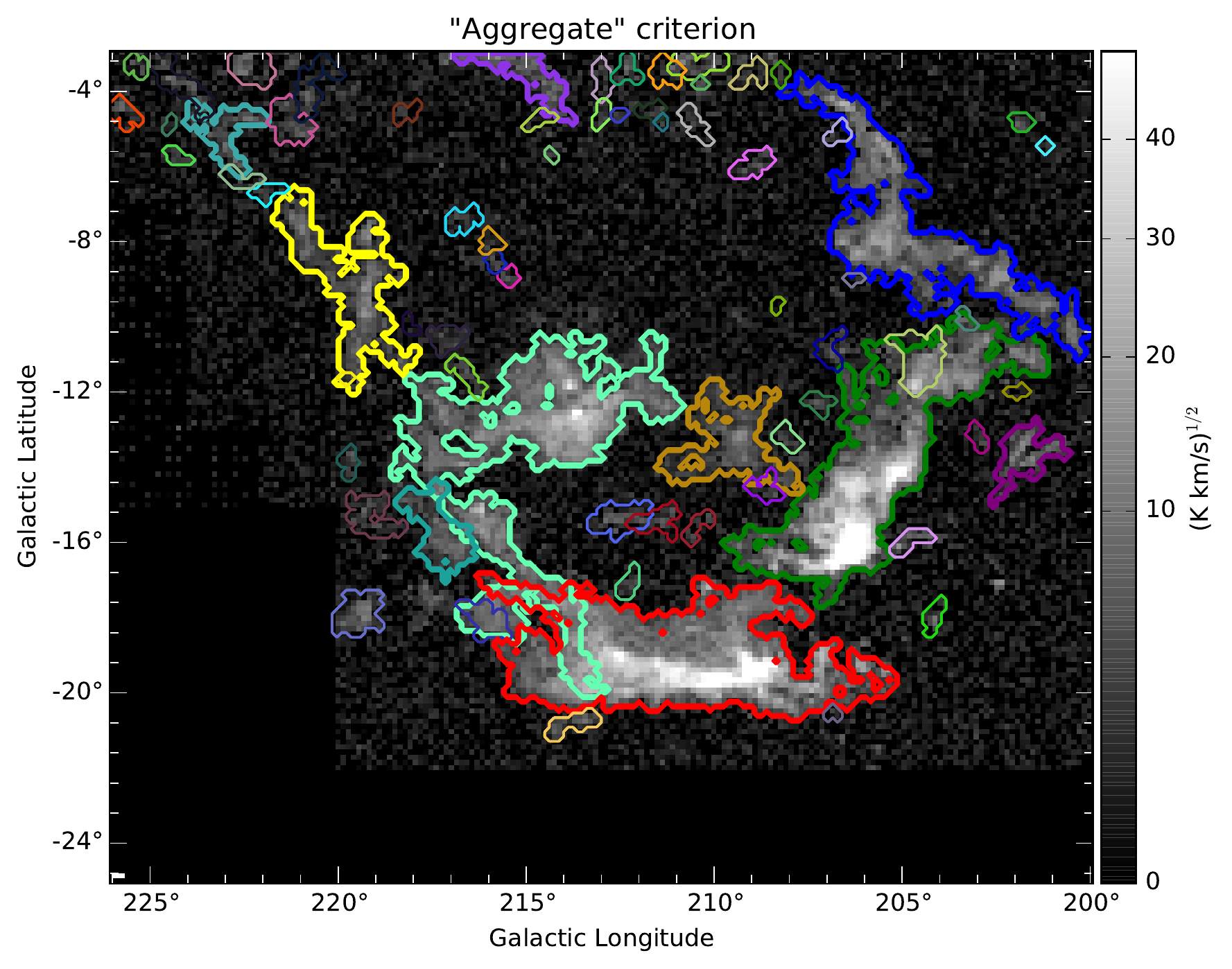}}
\end{center}
\caption{The Orion-Monoceros complex square root of the integrated intensity maps. Every different contour color represents a single cloud of the complex identified by \texttt{SCIMES}, through the volume, flux, and the aggregated criteria respectively, without including distance information. The contours use the same color scheme of figure~\ref{F:orion_dendro_nodist}.}
\label{F:orion_cont_nodist}
\end{figure}

\begin{figure*}
{\includegraphics[width=0.85\textwidth]{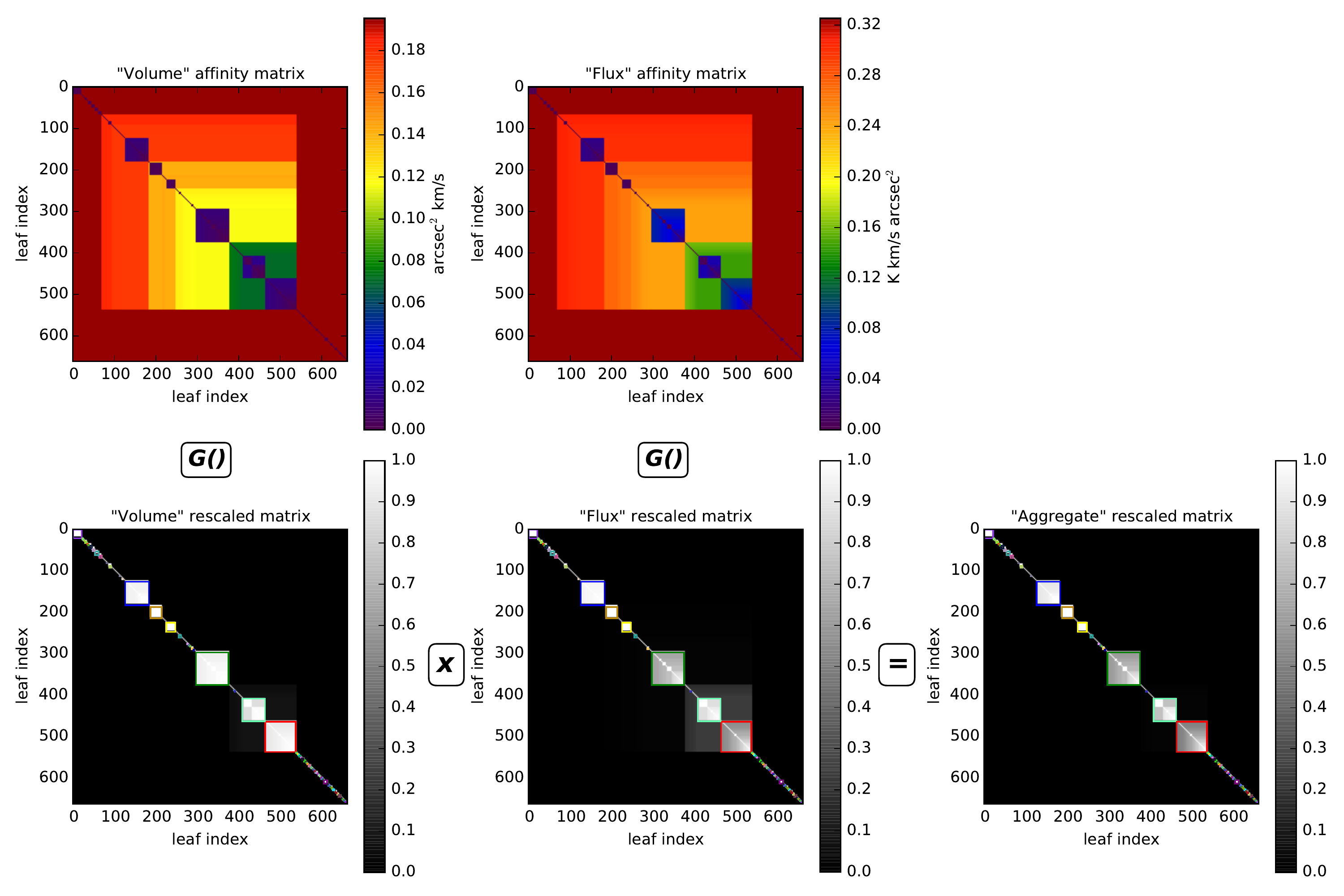}}
\caption{Similarity matrices (as in figure~\ref{F:orion_affmats}) obtained from the Orion-Monoceros complex dendrogram without including distance information.}.
\label{F:orion_affmats_nodist}
\end{figure*}

\begin{figure*}
{\includegraphics[width=0.75\textwidth]{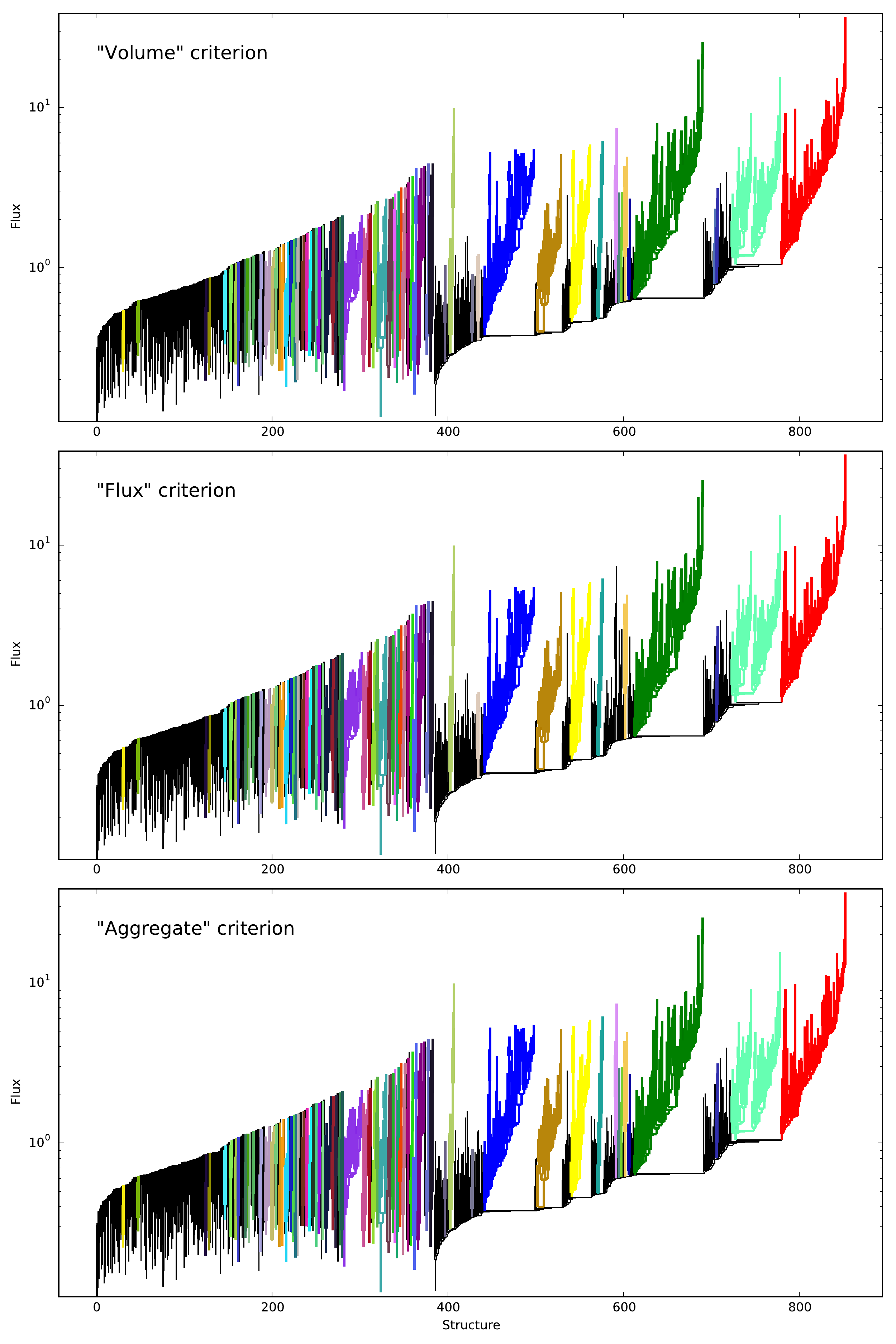}}
\caption{Dendrogram of the Orion-Monoceros complex obtained using the same parameters as in figure~\ref{F:orion_affmats_nodist} through (from the top to the bottom) the ``volume'', ``flux'' and ``aggregate'' criteria, respectively, without including distance information. Every color region outlines structures belonging to a certain cloud as segmented by SCIMES.}
\label{F:orion_dendro_nodist}
\end{figure*}

%=====================================================================
Robust distance estimations are rarely available especially for Galactic studies. In this Appendix, we show, therefore, how the Orion-Monoceros data segmentation changes without providing distance information to \texttt{SCIMES}. In this case, we need to consider the ``volume'' criterion as in equation~\ref{E:volume} but measured in arcsec$^2$\,km/s. The ``luminosity'' criterion is effectively turned into a ``flux'' criterion (equation~\ref{E:flux}) measured in K\,km/s\,arcsec$^2$. Figures~\ref{F:orion_cont_nodist}-\ref{F:orion_dendro_nodist} show the results of the segmentation in terms of contoured integrated intensity maps, affinity matrices, and clustered dendrograms, as provided by volume, flux and aggregate criteria. \texttt{SCIMES} selected $\sigma_S=0.04$\,arcsec$^2$\,km/s for rescaling of the volume matrix, and $\sigma_S=0.12$K\,km/s\,arcsec$^2$ for the flux matrix. Further, the code guessed $k_g=\{71,71,72\}$ number of clusters for the three criteria, respectively. The silhouette analysis pointed out to more appropriate $k_s=\{74,64,69\}$ clusters. The silhouette values for the clustering criteria are sil $=\{0.98,0.92,0.96\}$. The final cluster cleaning eliminates 3 clusters from the final count of volume-criterion based segmentation, leaving the number of cluster found by the algorithm equal to 71. 

One of the difference we note with respect to the distance-based decomposition is that the affinity matrices from both volume and flux are more alike: the blocks corresponding to the objects identified by \texttt{SCIMES} are well defined in both the matrices. In the same way, the silhouette calculated from the flux decomposition in larger than the luminosity based one. Dendrogram leaves form well separated clusters and all criteria provide quite accurate cloud segmentations. The main difference with respect to the decomposition generated attributing distances to the dendrogram structures is that Monoceros and NGC2149 are identified as a single entity. However these clouds appear spatially separated, having different orientation and morphology. Indeed, by turning the ``volume'' criterion into an ``area'' criterion (by using $A=\pi R^2$ instead of $V=\pi R^2\sigma_v$ to weight the graph edges) \texttt{SCIMES} separates this structure into two different clouds. Nevertheless, Monoceros and NGC2149 have similar velocity dispersions ($\sigma_v=1.6, 1.3$\,km/s, respectively) and centroid velocity ($V_{\textrm{cen}}=12.0, 12.7$\,km/s, respectively). These structures are coherent in velocity and they have strong affinity in the velocity dispersion space. The high affinity in velocity dispersion washes out the lower affinity introduced by the area and makes \texttt{SCIMES} to merge Monoceros and NGC2149 when the volume criterion (without distance information) is used. Instead, through the distances reported in Table 2 of \cite{wilson05} Monoceros is, in proportion to the units used, two times larger and the affinity with NGC2149 is much lower. In this case, spatial affinities dominate the velocity dispersion ones and the two objects are separated. Nevertheless, NGC2149 distance is not well determined. \cite{wilson05} attributed it by averaging the distances to Orion A and the Southern Filament, since background stars are not present for this cloud. We consider, therefore, that an association of NGC2149 with Monoceros is possible, given the high velocity coherence of the two objects.         

%=====================================================================
% END DOCUMENT
%=====================================================================

\bsp % ``This paper has been produced using the ...''

\label{lastpage}

\end{document}